\documentstyle[12pt,aaspp4]{article}
\def\Msunyr{\Msun ~\rm yr^{-1}}
\def\Mu5{\dot{\cal{M}}}
\def\EE#1{\times 10^{#1}}
\def\ccm{\rm ~cm^{-3}}
\def\kms{\rm ~km~s^{-1}}
\def\ergs{\rm ~erg~s^{-1}}
\def\etal{\rm et al.~}
\def\Ha{{\rm H}\alpha}

\def\Msunyr{~{\rm M}_\odot~yr^{-1}}
\def\Mdot{\dot M}

\def\lsim{\!\!\!\phantom{\le}\smash{\buildrel{}\over
  {\lower2.5dd\hbox{$\buildrel{\lower2dd\hbox{$\displaystyle<$}}\over
                               \sim$}}}\,\,}
\def\gsim{\!\!\!\phantom{\ge}\smash{\buildrel{}\over
  {\lower2.5dd\hbox{$\buildrel{\lower2dd\hbox{$\displaystyle>$}}\over
                               \sim$}}}\,\,}
\begin{document} 
\title{RADIO EMISSION AND PARTICLE ACCELERATION IN 
SN 1993J}
\author{Claes Fransson \altaffilmark{1} 
}
\centerline{and}
\author{Claes-Ingvar Bj\"ornsson \altaffilmark{1}}
\altaffiltext{1} {Stockholm Observatory, SE-133 36 Saltsj\"obaden, 
Sweden}
\authoremail{claes@astro.su.se, bjornsson@astro.su.se}    
\begin{abstract}
The radio light curves of SN 1993J are discussed. We find that a fit to the
individual spectra by a synchrotron spectrum, suppressed by
external free-free absorption and synchrotron self-absorption gives a superior fit to  models based on pure free-free absorption. A standard $r^{-2}$ circumstellar 
medium is assumed, and found to be adequate.
From  the flux and cut-off wavelength,  the magnetic field in the synchrotron
emitting region behind the shock is determined to $B
\approx 64~(R_{\rm s}/10^{15} {\rm ~
cm})^{-1}$ G.   The strength of the field argues strongly for turbulent
amplification behind the shock.
The ratio of the magnetic and thermal energy density behind the shock
is $\sim 0.14$.
Synchrotron losses dominate the cooling of the electrons, while
inverse Compton losses due to photospheric photons are less important.
For most of the time also Coulomb cooling 
affects the spectrum.
A model, where a constant fraction of the shocked, thermal electrons are injected and
accelerated, and subsequently lose their energy due to synchrotron losses,
reproduces the observed evolution of the flux and number of relativistic 
electrons well. The injected electron spectrum has $dn/d\gamma \propto
\gamma^{-2.1}$, consistent with diffusive shock acceleration. The injected 
number density of relativistic electrons  
scales with the thermal electron energy density, $\propto \rho V^2$,
rather than the density, $\rho$. 
The evolution of the flux is strongly connected to the deceleration of
the shock wave. 
The total energy density of the relativistic electrons, if extrapolated
to $\gamma \sim 1$, is $\sim
5\EE{-4}$ of the thermal energy density.
The free-free absorption required is consistent
with previous calculations of the circumstellar temperature of SN
1993J, $T_{\rm e} ~\sim (2-10)\EE5$ K, which failed in explaining the 
radio light curves by pure free-free absorption. 
Implications for the injection 
of the relativistic electrons, and the
relative importance of free-free absorption,  Razin suppression, and 
the synchrotron self-absorption
effect for other
supernovae are also briefly discussed. It is argued that especially
the expansion velocity, both directly and through the temperature, 
is important for determining the relative
importance of the free-free and synchrotron self-absorption.  Some
guidelines for the modeling and interpretation of VLBI observations are 
also given. 
\end{abstract}


\section{INTRODUCTION}
\label{sec-intro}
The radio emission from young supernovae provides an important probe
of the circumstellar environment of the supernova (Chevalier 1990,
Fransson 1994). In addition, the time-dependence of the physical conditions 
makes it a unique 
laboratory for studies of relativistic particle acceleration. 

For the first radio supernovae to be well observed, SN 1979C and SN
1980K (Weiler et al. 1986) a standard model to explain their radio light
curves was soon developed (Chevalier 1982b). The basic feature of this
model is  external free-free absorption in a circumstellar medium of
the synchrotron emission, arising close to the supernova shock
wave. Because of the expansion, the optical depth decreases as
$\tau_{\rm ff} \propto \lambda^2~ R_{\rm s}^{-3} \propto   \lambda^2~
t^{-3}$, if the shock velocity is constant. The intrinsic emission
from the shock, i.e. the injection efficiency of relativistic
particles, as well as the magnetic field, were assumed to scale with
the thermal energy density behind the shock.  This and alternative
scalings have been discussed by Chevalier (1996). This model has been
successful in explaining the observations of 'standard' Type II's
(Chevalier 1984), and with a knowledge of the temperature the mass
loss rate of the progenitor can be determined. Calculations of the
temperature (Lundqvist \& Fransson 1988) show  that $T_{\rm e} \approx
10^5 - 10^6$ K, and mass loss rates of $3\EE{-5}$ and $1.2\EE{-4}
\Msunyr$  were derived for SN 1980K and SN 1979C, respectively. The
model has since been applied to a number of other supernovae, although
the observational data have for most of these been less complete.

With the recent Type IIb SN 1993J the most complete data set of any
supernova have become available (Van Dyk \etal 1994; Pooley \& Green
1993). When the free-free absorption model was applied to this it was,
however, found that only a poor fit was obtained (Lundqvist 1994;
Fransson, Lundqvist, \& Chevalier 1996,  in the following FLC96; Van
Dyk \etal 1994).  An improved, although still unsatisfactory, fit was
obtained if the external density instead of $\rho \propto r^{-2}$
followed $\rho \propto r^{-d}$ where $d =1.5 - 1.7$.  The reason for
this flatter profile was not clear, although speculations about
variations in the mass loss rate have been proposed. Even more
complex models  could, not surprisingly, improve the fit, but at
the expense of a larger number of free parameters without obvious
physical meaning.  A serious problem, pointed out by FLC96, was that
the temperature in the circumstellar medium was $\gsim 2\EE5$ K,
resulting in a much smaller free-free optical depth than necessary to
produce the required absorption. 

In this paper we discuss  alternative explanations to the free-free
absorption model, and show that 
the radio spectra of SN 1993J are very well explained by a 
combination of synchrotron  self-absorption and external free-free
absorption, explaining the anomalous behavior of   SN 1993J, as compared to other radio
supernovae. Synchrotron self-absorption has been proposed to be
responsible for the low frequency cutoff in  Type Ib supernovae
(Shklovskii 1985, Slysh 1990), and independently from the present 
analysis has been discussed by Chevalier (1998). Chevalier \& Dwarkadas
(1995) have proposed that the turn on of the early radio emission  of
SN 1987A was determined by synchrotron self-absorption.  Several of
the mechanisms below were discussed by Chevalier (1982b), who,
however, for good reasons discarded them for SN 1979C and SN 1980K. 

In section \ref{sec-analys} we discuss the synchrotron emissivity at a
finite density of thermal electrons, including the Razin-Tsytovich
effect, and in section 3 we apply this analysis to 
the spectra of SN 1993J. We first use simple parameterized models with
constant spectral index and discuss the resulting fits for models with
the three different suppression mechanisms, i.e., free-free absorption, synchrotron 
self-absorption, and the Razin-Tsytovich effect. 
After this we discuss a
self-consistent model where the spectral index is determined by the
various energy loss mechanisms. In section 4 we then discuss the 
results  of this
analysis, and their implications for the magnetic field, 
relativistic particle acceleration and circumstellar temperature. 
Section 5 contains models for the full light curves, while in section 6 we discuss 
the relative importance of the Razin-Tsytovich effect, synchrotron self-absorption and
free-free absorption. In section 7 we make some comments on the analysis 
of VLBI observations of supernovae. The conclusions are
summarized in section 8.

\section{ABSORPTION AND SUPPRESSION MECHANISMS FOR THE SYNCHROTRON EMISSION}
\label{sec-analys}
A main advantage in analyzing the individual radio spectra for the
various observed dates is that they are only a function of the
spectral index of the synchrotron emission, the normalization of the
spectrum, and a one-parameter or at most two-parameter specified
function,  describing the effect of the external or emitting
medium. In contrast, an analysis based on the light curves of the
different wavelength bands involves uncertain assumptions about the
particle acceleration mechanism (e.g., Chevalier 1996). From the time
variation of the parameters involved in the fitting of the individual
spectra we can instead hope to gain some understanding of these
uncertain, but extremely important, properties of the particle
acceleration. 

For a homogeneous shell with a synchrotron
optical depth $\tau_{\rm s}(t)$, and with an external
free-free medium, parameterized by the  optical depth $\tau_{\rm
ff}(t)$ at a wavelength $\lambda = 1$ cm,  Gaunt
factor $g_{\rm ff}(\lambda)$, and impact
parameter $s$, the spectrum is given by 
\begin{equation}
F_\nu(\lambda) = 2 \pi \left({R_{\rm s} \over D}\right)^2 S_{\nu}
  ~\int_0^1 ~[1 - e^{-\tau_{\rm
s}(s,t)}]~  e^{-\tau_{\rm ff}(s,t)~g_{\rm
ff}(\lambda) ~\lambda^{2}} ~s~ ds
\label{eqssa}
\end{equation}
where 
\begin{equation}
 \tau_{\rm s}(s,t)={\tau_{\rm s}(0,t)   \over (1-s^2)^{1/2}},
\label{eqssa1}
\end{equation}
and
\begin{equation}
\tau_{\rm ff}(s,t) = \int_{R_s}^\infty {\kappa_{\rm ff}(r)  n_{\rm
e}(r)^2 r \over (r^2-s^2 R_s^2)^{1/2}} dr.
\label{eqssa2}
\end{equation}
Here, $\kappa_{\rm ff}$ is the free-free
 opacity at 1 cm and  $n_{\rm e}$ the thermal electron density. 
$S_{\nu}$ is the source
function, $S_{\nu} = j_{\nu}/
\kappa_{\nu}$, with $\kappa_{\nu}$ the synchrotron self-absorption opacity and
$ j_{\nu}$ the emissivity. Finally, $R_s$ and $D$
are  the radius of the emitting shell and the distance to the
supernova, respectively. We assume that the supernova ejecta are opaque
to the radio emission, as is most likely the case.  

In the analysis in
section \ref{sec-onecomp}, we assume that the column density of electrons have a
constant power-law, giving a synchrotron spectrum with index
$\alpha$. If we further replace equation (\ref{eqssa}) by an
average over the disk we then get
\begin{equation}
F_\nu(\lambda) = S'(t)~\lambda^{-5/2}~[1 - e^{-\bar{\tau}_{\rm
s}(t)~\lambda^{(\alpha+5/2)}}]~  e^{-\bar{\tau}_{\rm ff}(t)~g_{\rm
ff}(\lambda) ~\lambda^{2}} 
\label{eqssab}
\end{equation}
where the parameter
$S'(t)$ is related to $S_\nu$ at 1 cm by $S' = \pi R_s^2 ~S_\nu(1{~\rm cm})/D^2$. 
The average optical depth over the disk, $\bar{\tau}_{\rm s} $, is given by
$\bar{\tau}_{\rm s} \approx 2 
\kappa_{\rm s}\Delta R_{\rm em}$, where $\Delta
R_{\rm em}$ is the thickness of the emitting shell. In the optically thin limit we have
\begin{equation}
 F_{\nu}(\lambda) \approx 2 \pi \left({R_{\rm s} \over D}\right)^2 j_{\nu}(t)
\Delta R_{\rm em} ~e^{-\bar{\tau}_{\rm ff}(t)~g_{\rm ff}(\lambda)
 ~\lambda^{2}} .
\label{eqa} 
\end{equation}

In the general case of a plasma with density $n_{\rm e}$, magnetic field
$B$, and relativistic electron distribution $N(\gamma)$ (integrated
along the line-of-sight, and assumed to have an isotropic pitch angle
distribution),  the 
line-of-sight integrated emissivity is given by (Tsytovich 1951; Razin 1960;
Pacholczyk 1970)
\begin{equation}
 j_\nu \Delta R_{\rm em}= 1.87\EE{-23}~ B~ \sin \theta ~\int_1^\infty ~N(\gamma) f({\nu_{\rm p} \gamma \over \nu})~F[{\nu \over \nu_{\rm
c}  f({\nu_{\rm p} \gamma \over \nu})^3}]~ d\gamma ~ 
\label{eql} 
\end{equation} 
where $\theta$ is the pitch angle, $\nu_{\rm p} = 8.98\EE3 n_{\rm
 e}^{1/2} $ Hz is the plasma frequency, and $\nu_{\rm c} = \nu_0 B
 \sin \theta  ~ \gamma^2$  is the critical frequency, where $\nu_0 = 4.20\EE6
 $ Hz.  $F(x)$ is given
 by 
\begin{equation}
 F(x) = x~\int_x^\infty ~K_{5/3}(z) ~dz
\label{eqd} 
\end{equation}
where $K_{5/3}(z)$ is the modified Bessel function of order- 5/3.  The
function $f(x) = (1 + x^2)^{-1/2}$
represents the modification due to plasma effects.  Similarly, the
synchrotron self-absorption optical depth is given by 
\begin{equation}
 \tau_\nu = 1.02\EE{4}~ {B~ \sin \theta \over \nu^2} ~\int_1^\infty
 ~\gamma^2 ~ {d  \over d \gamma}~ \left[ \gamma^{-2}~  N(\gamma) \right] f({\nu_{\rm p} \gamma \over \nu})~ F[{\nu \over
\nu_{\rm c} f({\nu_{\rm p} \gamma \over \nu})^3}]~  d\gamma ~ 
\label{eqla} 
\end{equation}

For a constant power law distribution of electrons, $N(\gamma) \equiv N_{\rm
rel}~\gamma^{-p}$, one finds
\begin{equation}
 j_\nu \Delta R_{\rm em}= 9.33\EE{-24}~
\left(1.40\EE{-4}~\lambda\right)^{\alpha}~\left(B~ \sin \theta
\right)^{(\alpha+1)}~N_{\rm rel}~\int_0^\infty ~g(x)~F[{x \over
g(x)^3}]~ x^{(\alpha-1)} ~dx ~ 
\label{eqlb} 
\end{equation} 
where 
\begin{equation}
g(x) \equiv  (1 + {\lambda \over \lambda_{\rm R}~x})^{-1/2},
\label{eqm1} 
\end{equation}
and $\alpha = (p -1)/2$.  The critical wavelength, $\lambda_{\rm R}$,
above which the emission is suppressed is given by 
\begin{equation}
 \lambda_{\rm R} = {3 \over 4}~{B \sin \theta \over e~n_{\rm e}} =
\left({n_{\rm e} \over 1.56\EE9 \ccm}\right)^{-1}~B~\sin \theta ~~~\rm
cm.
\label{eqc} 
\end{equation}

\begin{figure}  
\plotone{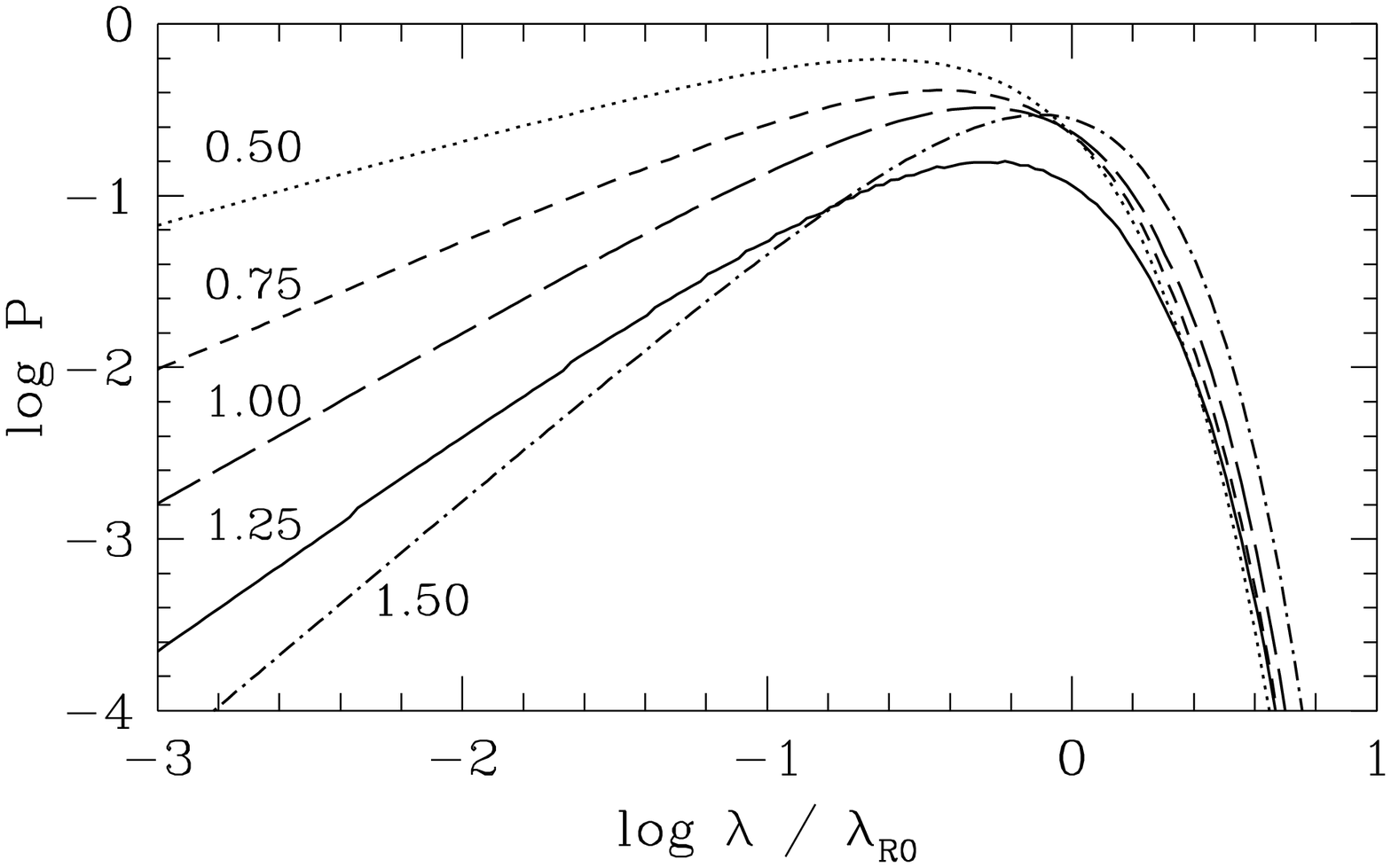}
\caption{Angle averaged synchrotron spectra, including the Razin effect, 
for different  power law indices, $\alpha$. The Razin wavelength is given by 
$\lambda_{\rm R0} = 1.56\EE9 B / n_{\rm e}$ cm.}
\label{fig1}
\end{figure}

Equation (\ref{eql}) contains the pitch angle between the magnetic
field and the electron velocity. Microscopically, the magnetic field
may be random because of turbulence behind the shock. In the
integrated flux, as measured by the Very Large Array (VLA), one observes even a well ordered
field from 
different directions, because of the spherical  geometry of the
shock. We therefore average equation (\ref{eql}) over all angles
numerically. We also assume that the magnetic field and density are
constant over the emitting volume. Further, to  isolate the wavelength
dependence from the variation in the normalization, we split the
flux into one wavelength independent amplitude, $A(t)$, and one
spectral function  $P(\lambda / \lambda_{\rm R0}, \alpha)$,
\begin{equation}
F_\nu(\lambda)= A(t)~P(\lambda / \lambda_{\rm R0}, \alpha).
\label{eqsa} 
\end{equation}
Here, 
\begin{eqnarray}
A(t) = 5.86\EE{-23}~ \left({n_{\rm e} \over 2.19\EE{5}
\ccm}\right)^{-\alpha}~B^{2 \alpha+1}~N_{\rm rel} \left({R_{\rm s}
\over D}\right)^2 
\label{eqr} 
\end{eqnarray}
and $\lambda_{\rm R0} = \lambda_{\rm
R}(\theta=\pi/2)$. The spectral function $P(y, \alpha)$ is given by
\begin{equation}
P(y, \alpha)= {y^{\alpha} \over 2}~ ~\int_0^\pi \sin
\theta^{(\alpha+1)} \int_0^\infty~(1 + {y\over \sin \theta~x})^{-1/2}~
F[x (1 + {y \over \sin \theta~x})^{3/2}]~ x^{(\alpha-1)} ~dx ~ d \cos
\theta ~.
\label{eqra} 
\end{equation} 
$P(y, \alpha)$ is shown in figure \ref {fig1} for several values
of $\alpha$.  The averaged and non-averaged spectrum only differ
marginally from each other. The averaged spectrum has a slightly
shallower slope close to the maximum.


At wavelengths longer than $\lambda_{\rm R0}$ the emission is severely
suppressed. This effect has been referred to as alternatively the Razin-Tsytovich effect
or simply the Razin effect. In the following we will for simplicity use the latter
terminology. The vacuum case is obtained for
$\lambda_{\rm R0}
\rightarrow \infty$ when $f = 1$.  For wavelengths longer than
$\lambda_{\rm R0}$ the pitch angle averaged spectrum falls off as 
\begin{equation}
F_\nu \propto N_{\rm rel}~B^{\alpha+1}~\lambda^{\alpha}~\left(\lambda
\over  \lambda_{\rm R0} \right)^{(\alpha-1/2)}~e^{-{3^{3/2} \over
2^{1/2}}~ {\lambda \over \lambda_{\rm R0}}}
\label{eqj} 
\end{equation}
(Simon 1969).  The cutoff  due to the Razin effect is therefore
less steep than that due to free-free absorption. This is important
for the shape of the spectrum at long wavelengths.

The relevant values of $n_{\rm e}$ and $B$ depend on the particle
acceleration site.  If  this is in the shocked circumstellar gas,
$n_{\rm e}$ will be a factor of four higher than the electron density
in the wind.  In FLC96 and Fransson (1998) the X-ray emission of SN
1993J at early and late time is modeled. To reproduce the observed
X-ray flux a mass loss rate of $\Mdot \approx 5\EE{-5} \Msunyr$ is
required, for a wind velocity of $u_{\rm w} = 10\kms$. With a factor
four  compression, and $X({\rm He})/X({\rm H}) = 0.3$ (e.g., Shigeyama
\etal 1994), the electron density immediately behind the circumstellar
shock is 
\begin{eqnarray}
n_{\rm e}  =&&
4.36\EE8~ \dot{\cal{M}}~ \left({R_{\rm s} \over 10^{15} ~{\rm
cm}}\right)^{-2} \cr =&& 1.46\EE8 ~\dot{\cal{M}}~ \left({V \over 2\EE4
\kms} \right)^{-2}  \left({ t \over 10~ {\rm days}}\right)^{-2}~\rm
cm^{-3}
\label{eqac} 
\end{eqnarray}
The similarity solution by Chevalier (1982a)  shows that the density
behind the circumstellar shock is fairly constant (see also figure 3
in FLC96), and we will use equation (\ref{eqac}) throughout the shocked
region.  We have here for future convenience defined a normalized mass
loss, $\dot{\cal{M}}$, rate by 
\begin{equation}
\dot{\cal{M}} \equiv \left({\Mdot\over 5\EE{-5} \Msunyr}\right)
\left({ u_{\rm w} \over 10 \kms}\right)^{-1}
\label{eqaca} 
\end{equation}
In FLC96 it is argued that $V \approx 2\EE4 \kms$ at $\sim 10$ days.
VLBI observations (Marcaide \etal 1995a,b, 1997; Bartel \etal 1994)
give an expansion velocity for the radio emitting plasma  of $\sim
(1.8-2.4)\EE4 \kms$  during the first $\sim 100$ days.
 Consequently, we scale our results to $V = 2\EE4
\kms$. The motivation for this and the evolution of the velocity is 
discussed further in section \ref{sec-velev}.

With this density $\lambda_{\rm R0}$ is given by 
\begin{equation}
 \lambda_{\rm R0} = 10.7 ~\Mu5\left({B \over 1~ {\rm
G}}\right)\left({V \over 2\EE4 \kms}\right)^{2}  \left({ t \over 10~
{\rm days}}\right)^{2}~\rm cm.
\label{eqh}
\end{equation}
The magnetic field in the emitting region therefore has to  be
less than $\sim 1$ Gauss, or alternatively, $\Mdot /u_{\rm w}$ large,
for  the Razin effect to be important.

\section{APPLICATION TO SN 1993J}
\label{sec-93j}
Although radio observations are available for a large number of
supernovae (e.g., Weiler \etal 1998 for a review), SN 1993J is unique
in terms of both frequency of observations, small errors in the
observed fluxes, and the large wavelength coverage. It is therefore by
far the object best suited for a detailed analysis. Observations have
been published by Van Dyk \etal (1994) and Weiler \etal (1998)  for
1.33 -- 21 cm from 5 -- 1300 days, and by Pooley \& Green (1993) at 2
cm from 8 -- 114 days. In addition, there is one set of VLA
observations at 923 days from 1.3 -- 90 cm (Montes \etal  1995). A few
mm observations with IRAM and OVRI also exist  (Radford 1993; Phillips
\& Kulkarni 1993a, b). The fluxes of the mm observations do,  however, 
not agree at 
similar epochs, and it is difficult to judge the accuracy of these
observations. For the modeling of the spectra  in this paper we
use the VLA observations by Van Dyk \etal and Weiler \etal, as well as
the  day 923 observation by Montes \etal In section
\ref{sec-lum} we also include the Ryle telescope 
observations by Pooley \& Green
(1993), as well as the VLA dates having only a partial spectral
coverage, in particular before $\sim 10$ days. 

Our strategy is first to investigate the simplest models, based
on a pure power law  electron and synchrotron spectrum, studying one
suppression mechanism at a time. By analyzing   the results of this,
in particular the consistency of these models,
we test their validity as well as
explore the likely range of the   relevant physical parameters, in
particular the magnetic field.  However, as a result of the failure of
these  simple models we are in section \ref{sec-selfc} led to a more detailed
analysis, taking all the relevant physical processes  into account. 

\subsection{Constant spectral index fits}
\subsubsection{One component models}
\label{sec-onecomp}
We will now use equation (\ref{eqssab}), assuming here a constant spectral index $\alpha = (p-1)/2$, in the three limiting cases: 1) free-free absorption,
$\tau_{\rm s} = 0$, $\tau_{\rm ff} \neq 0$, $\lambda \ll \lambda_{\rm
R0}$; 2) synchrotron-self absorption, $\tau_{\rm s} \neq 0$, $\tau_{\rm ff} = 0$, $\lambda
\ll \lambda_{\rm R0}$; 3) Razin suppression, $\tau_{\rm s}
= 0$, $\tau_{\rm ff} = 0$, $\lambda \approx \lambda_{\rm R0}$. 
In the free-free and Razin cases we use the optically thin
expression, equation (\ref{eqa}).  In all three cases 
there are  three parameters
characterizing the spectrum: 1) the overall normalization of the
synchrotron spectrum, 2) the power law index of the radiation ($\alpha$),
3) the optical depth to free-free absorption ($\tau_{\rm ff}$), synchrotron self-absorption ($\bar{\tau}_{\rm s}$),  or the Razin cutoff wavelength
($\lambda_{\rm R0}$). For each date we fit the observed spectrum with
the set of parameters that minimizes $\chi^2$.  The value of $\alpha$
is determined by minimizing the scatter of the overall light curves;
in the Razin case, especially the spectral region where $\lambda \ll  \lambda_{\rm
R0}$, and for the free-free and synchrotron self-absorption cases the spectral region
where $\tau(\lambda) \ll 1$. As we discuss below there are reasons to
expect that, because of cooling, $\alpha$  varies with time and frequency. 
This will be discussed in detail in section \ref{sec-lum}.

\begin{figure}  
\plotone{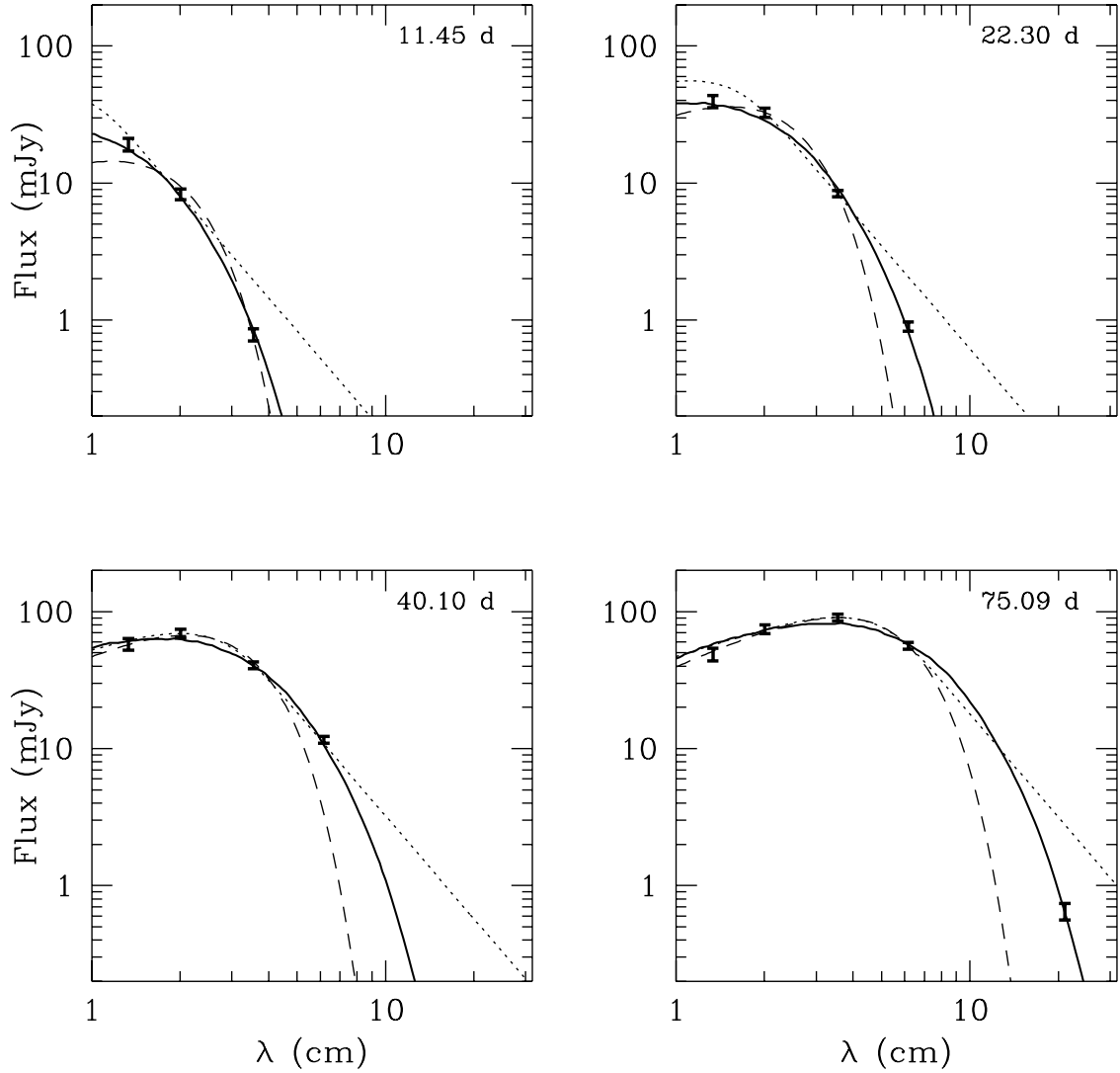}
\caption{Observed VLA spectra at four epochs, from Van Dyk \etal (1994), 
together with fits based
on a Razin suppressed synchrotron spectrum (solid line), external free-free 
absorption (dashed line), and  synchrotron self-absorption (dotted line).}
\label{fig2}
\end{figure}

In figure \ref{fig2}  we show the observed fluxes at 1.3, 2.0, 3.6,
6.2, and 21 cm at 11.5, 22.3, 40.1, and 75.1 days. These dates were
picked to cover well the exponentially damped part of the spectrum. 
 The solid line represents the  best fit Razin spectrum,
while the dashed line is the best fit free-free absorption spectrum and
the dotted line shows the same for the synchrotron self-absorption case. In
the Razin case $\alpha = 1.25 \pm 0.1$ gives the best fit. 
In the free-free case $\alpha = 0.8$ is indicated, while the best fit
in the synchrotron self-absorption case is obtained for $\alpha =
0.7$. The reason for this difference in $\alpha$ is that the Razin  effect
affects the spectrum well below $\lambda_{\rm R0}$, and flattens the overall
spectrum. The free-free absorption sets in more abruptly, and has less
an effect at short wavelengths (cf.  equations \ref{eqa} and
\ref{eqj}). 

From figure \ref{fig2} it is obvious that the Razin spectrum gives a
superior fit to the observations compared to  free-free or synchrotron
self-absorption. Only when all wavelengths are close to the
non-suppressed level, as at 40.1 days, do   free-free and
synchrotron self-absorption  provide  acceptable fits.  Otherwise the
free-free spectrum underestimates the flux, while the synchrotron
self-absorption overestimates above the wavelength cutoff. 
At $t \gsim 200$ days all models give reasonable fits because
of the decreasing importance of both the Razin effect, synchrotron
self-absorption, and free-free absorption at $\lambda \lsim 20$ cm. 

\begin{figure}  
\plotone{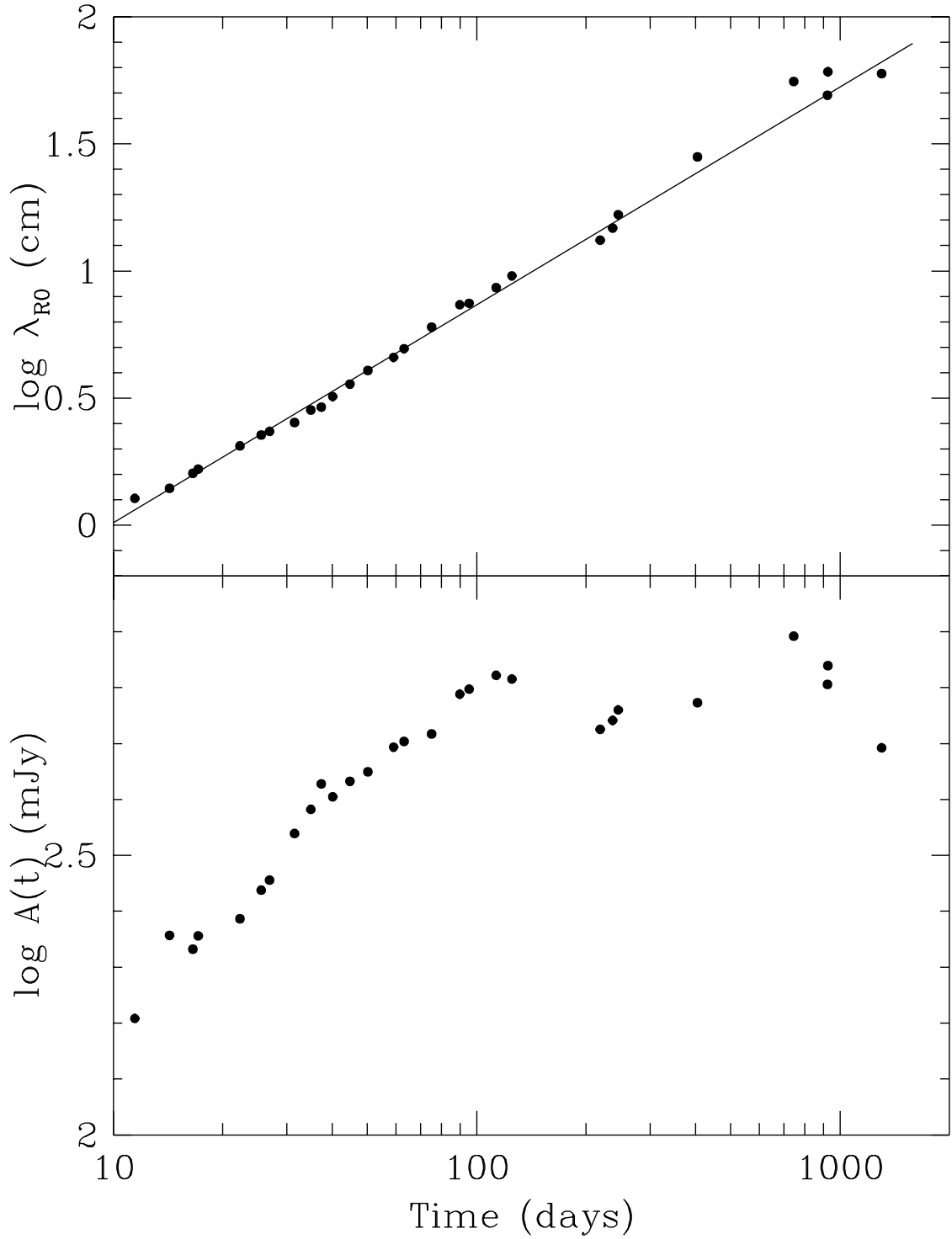}
\caption{Derived values of $\lambda_{\rm R0} \propto B / n_{\rm e} $ 
and $A(t)$ from the
fits to the Razin model. The solid
line gives a best linear fit to the $\log B/n_{\rm e}$ versus $\log t$ relation.
}
\label{fig3}
\end{figure}

Because the Razin fit provides the best one component fit, we now
analyze the implications and consistency of this model in more detail.
 We have therefore calculated the best fit for all dates
 which have enough
frequency points (three for $t < 16$ days, four for $16 < t < 75$
days, and five for $75 {\rm ~days}< t$). The Razin wavelength, $\lambda_{\rm R0}$,
 and the
normalization parameter $A(t)$ (in mJy) (see equation \ref{eqsa}) 
are calculated for these dates and shown 
in figure \ref{fig3}.
During the whole period the Razin wavelength
shows a very smooth evolution. The solid line in figure \ref{fig3}
shows a simple power law fit to the evolution, determined by a least
squares fit.  Using equation (\ref{eqc}), the ratio $B/n_{\rm e}$  can
be fit with the function
\begin{equation}
{B \over n_{\rm e}} = 6.6\EE{-10}~\left({t \over 10 {\rm~ days}}
\right)^{0.86}
\label{eqe} 
\end{equation}
with an r.m.s. of only $\pm 0.013$ in the exponent. 
Inserting equation (\ref{eqac}) 
into equation (\ref{eqe}), the magnetic  field  can be obtained
for the same region as a function of time,
\begin{eqnarray}
B   \approx 9.6\EE{-2}~\Mu5~ \left({V \over 2\EE4 \kms} \right)^{-2}
\left({ t \over 10~ {\rm days}}\right)^{-1.14}~\rm G.
\label{eqq} 
\end{eqnarray}

To calculate the total number of relativistic electrons we eliminate
in equations (\ref {eqsa}) and (\ref {eqr}) the magnetic field and
electron density by using equations (\ref {eqq}) and (\ref{eqac})  and find
with $p=3.5$
\begin{eqnarray}
F_{\nu} =&& 
1.14\EE{-23}~\Mu5^{2.25} ~\left({V \over 2\EE4 \kms}
\right)^{-2.5}~N_{\rm rel}~ \cr &&  P({\lambda \over
\lambda_{\rm R0}})\left({t\over 10 ~{\rm days}}\right)^{0.5} \rm ~ mJy
\label{eqwk} 
\end{eqnarray}
The observed spectral evolution can for $t \lsim 100$ days be fitted by 
\begin{equation}
F_{\nu} = 1.6\EE{2}~P({\lambda \over \lambda_{\rm R0}})\left({t\over
10~{\rm days}}\right)^{0.62} ~\rm mJy.
\label{eqaz} 
\end{equation}
Therefore, 
\begin{eqnarray}
 N_{\rm rel} = 1.4\EE{25}~\Mu5^{-2.25}  ~\left({V \over 2\EE4 \kms}
\right)^{2.5}~\left({t\over 10 ~{\rm days}}\right)^{0.11} {\rm cm^{-2}}
\label{eqyk} 
\end{eqnarray} 
for $t \lsim 100$ days. 

As a consistency check of the assumption of $\tau_{\rm s} \ll 1$, 
from equation (\ref{eqla}) we can now calculate  the synchrotron
self-absorption optical depth with the   above values of $B$ and
$N_{\rm rel}$. One then finds that, due to the Razin suppression of
the opacity, the optical depth at $\lambda_{\rm R0}$ does not exceed 0.1 at any time. 

A fatal problem for the Razin model occurs when we estimate the
energy density of relativistic electrons, $u_{\rm rel}$.
We will be conservative here and just include electrons
with Lorentz factors responsible for the observed spectrum between 10 - 100 days,
i.e., wavelengths 1.3 -- 6.3 cm, giving $1.1\EE2 (t / 10 {\rm ~ days})^{0.57} \lsim
\gamma \lsim 2.4\EE2 (t / 10 {\rm ~ days})^{0.57}$. From equation
(\ref{eqyk}) and $p = 3.5$ we find that the energy per
unit area within this band is $\sim 4.7\EE{15} (t / 10 {\rm ~
days})^{-0.75} {\rm ~ergs ~cm^{-2}}$, which is only weakly dependent on $p$.  
As we show in section
\ref{sec-invcomp} (equation \ref{eqtcomp} with $L_{\rm bol} \approx 4\EE{42} ~(t/10
{~\rm days})^{-0.9} \ergs$), inverse Compton losses are in this case likely to be important 
down to $\gamma \approx 30$.  
This implies in the Razin case that the  
inverse Compton timescale during the first 100 days is  
$t_{\rm Comp}\sim 0.36~ (t / 10 {\rm ~
days})^{2.4} \lambda^{0.5}$ days. The thickness of the emitting region
is $\Delta R_{\rm em} \approx V t_{\rm Comp} /4$, or for $\lambda \sim
3$ cm
$\Delta R_{\rm em} \approx 2.5\EE{13}~ (t / 10 {\rm ~ days})^{2.4}$
cm. The energy density is then $u_{\rm rel} \approx 1.9\EE2 ~(t / 10
{\rm ~ days})^{-3.1}  {\rm ~ergs ~cm^{-3}}$. This can be compared to
the thermal energy density behind the shock, $u_{\rm
thermal} \approx  3.4\EE2 ~\Mu5~ (t / 10 {\rm ~ days})^{-2}  {\rm ~ergs
  ~cm^{-3}}$, which is only slightly higher than $u_{\rm rel}$.

Our estimate of $u_{\rm rel}$ is most
likely a serious underestimate, since we have  only
included the spectrum between wavelengths 1.3 -- 6.3 cm (already at 21 cm, 
$\gamma \sim 60 ~(t / 10 {\rm ~ days})^{0.57}$). In reality one should
extrapolate the electron spectrum down to $\gamma \sim 1$, 
which could increase the energy density by a factor of $ \gamma(\lambda = 1.3 ~{\rm
 cm})^{1.5} \gsim 300$. The exact factor depends on the detailed
cooling processes as well as the form of the spectrum close to
$\gamma \sim 1$, and is therefore model dependent. This will, however,
not alter our main conclusion that 
{\it the energy density of relativistic electrons in
this model is likely to exceed the energy density of thermal particles by
a large factor}, which is unrealistic in any model for the particle
acceleration, and rules out this model. Typical values of the
injection efficiency are instead $\lsim 10^{-2}$ (e.g., Chevalier
1982b). The main reason for this problem is the weak magnetic field and 
the steep power law index
derived from the Razin cutoff. Because $F_\nu \propto N_{\rm
rel}~B^{\alpha+1}$ this results in a very large value of $N_{\rm rel}$.

\subsubsection{Synchrotron self-absorption plus free-free external absorption}
\label{sec-ssaff}
The fact that the Razin model requires an unrealistically large
density of non-thermal electrons, due to the low magnetic field,
forces us to more complex models. The most natural one is then a
combination of synchrotron self-absorption and external free-free
absorption. The latter component   has the effect of suppressing the
too high flux at long wavelengths in the synchrotron  self-absorption
model. As we will find, this model also solves the efficiency problem above
by requiring a higher magnetic field, while at the same time giving a
free-free absorption consistent with earlier estimates in FLC96. 

To model the combined absorption we now use  equation (\ref{eqssa}),
integrated over the supernova disk, but still assuming a fixed spectral index. We then have $\tau_{\rm s}(1 ~{\rm cm})$ and
$\tau_{\rm ff}(1 ~{\rm cm})$ as free parameters, in addition to $\alpha$ and
$S'$.

For a given temperature, $T_{\rm e}(r)$,  and including the Gaunt
factor, $g_{\rm ff}$, the  free-free optical depth along  a ray with impact
parameter $s$ is given by 
\begin{equation} \tau_{\rm ff}(t,s) =
5.63\EE{53}~\dot{\cal{M}}^{2} \lambda^2~{(1+2X) \over (1+4X)}
\int_{R_s}^\infty  {(-2.78 +  \log[\lambda ~T_{\rm e}^{3/2}(r)]) \over
T_{\rm e}(r)^{3/2}~ r^3 ~(r^2-s^2 R_{\rm s}^2)^{1/2}} dr
\label{eqtauff1}
\end{equation}
where $X = X({\rm He})/X({\rm H})$. 
$\tau_{\rm ff}$ is therefore sensitive to the radial temperature profile, as
well as its dependence on time. Lundqvist \& Fransson (1988), as well
as FLC96, have discussed the determination of this extensively, and it
is found that at radii, $\lsim 1\EE{16}$ cm the gas cools rapidly from
$\gsim 1\EE6$ K to $\sim 2.5\EE5$ K, while at large radii the
temperature is  fairly constant  at $\sim (2-2.5)\EE5$ K. Therefore,
as the shock propagates outwards, the gas close to the shock will
initially have a decreasing temperature, while later it will be more
constant. This is seen clearly for SN 1993J in figure 11 of FLC96.  While this
behavior is probably quite generic, the details of the radial
variation depend strongly on the properties of the shock radiation
(see section \ref{sec-freefree}). 

In this paper we use a simple  parameterization of the temperature
profile, assumed to be constant with time, and given by
\begin{equation}
T_{\rm e}(r)= \max \left[ T_{15}~\left({10^{15}~{\rm cm} \over
r}\right)^\delta, ~2\EE5 {\rm~ K}\right] .
\label{eqtprof}
\end{equation}
We then vary the parameters $T_{15}$ and $\delta$ to get an optimum
fit of the spectra.  
Although indicative of the actual temperature, equation
(\ref{eqtprof}) should only be interpreted as a convenient fitting
formula.  The reason is that the fundamental parameter for the fitting
of the spectra is not  $T_{\rm e}(r)$, but rather $\tau_{\rm
ff}(t)$. As shown by equation (\ref{eqtauff1}), the same $\tau_{\rm
ff}(t)$ can be obtained by different $\dot{\cal{M}}$ and $T_{\rm
e}(r)$, as long as $\dot{\cal{M}}^{2} \int T_{\rm e}(r)^{-3/2} r^{-4}
dr \approx$  constant.  Different temperature profiles can therefore
yield the same result for $\tau_{\rm ff}(t)$. The value of $T_{\rm e}$
given by equation (\ref{eqtprof}) should therefore only be interpreted
as a weighted average. Also, the dependence of $T_{\rm e}$  on
$\dot{\cal{M}}$ should be kept in mind when interpreting the results. 

We have calculated the best fit for different values of $\alpha$,
$T_{15}$, and $\delta$, varying $ S'(t)$ and $\tau_{\rm s}(t)$ for
each date.  We find that good fits are obtained for  $\delta = 1.0$
and $T_{15} = 2.75\EE6$ K, with $\chi_\nu^2 = 1.4 - 1.5$. Larger and
smaller values of $\delta$ give considerably higher $\chi_\nu^2$. The
free-free absorption is only  well-determined by the spectra and light
curves later than $\sim 10$ days. 
We have tested different values of the spectral index, and
find that we get a best fit with $\alpha = 0.85$, or $p=2.7$. Because 
the quality of the fit is very similar to that in the more 
detailed model in figure \ref{fig8}, we do not show it here.

\begin{figure}  
\plotone{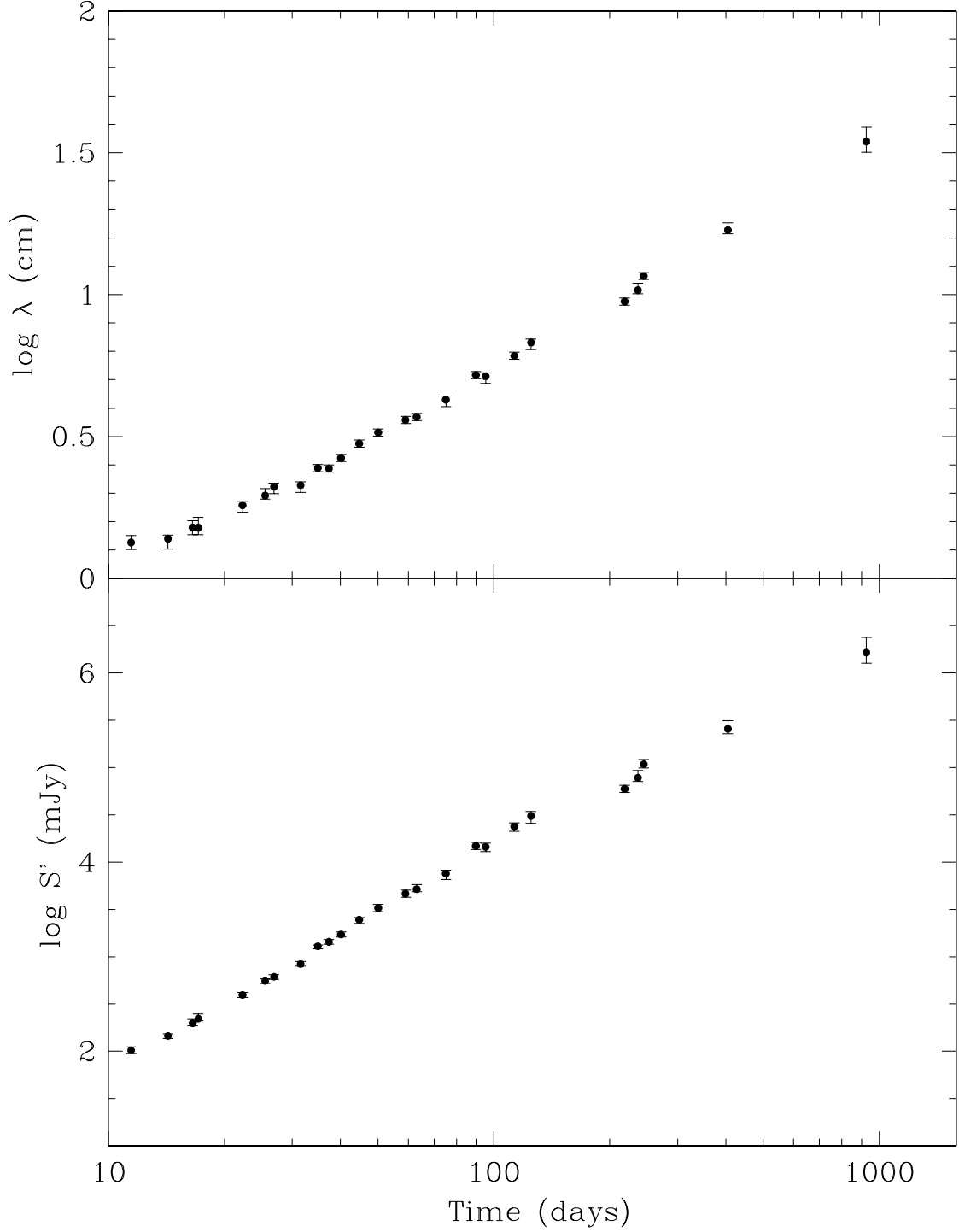}
\caption{Evolution of $\lambda_{\tau_{\rm ssa}=1}$ and the scaled source 
function $S'(t) = \pi S(t)~R(t)_s^2/D^2 $ 
for the synchrotron self-absorption model with external free-free absorption.}
\label{fig4}
\end{figure}

In figure \ref{fig4} we show the evolution of the wavelength at
which the  synchrotron self-absorption optical depth is unity,
$\lambda_{\tau_{\rm ssa}=1} = \tau_{\rm s}(t)^{-1/(5/2 + \alpha)}$,
and the scaled source function at 1 cm, $S'(t)$. A least squares fit to
the observations  for $10  < t < 100$ days gives 
\begin{equation}
\lambda_{\tau_{\rm ssa}=1} = \lambda_{10} ~\left({t \over 10{\rm
 ~days}}\right)^{q_\lambda} ,
\label{eqtaussa1}
\end{equation}
with $\lambda_{10} = 1.07$ cm, and $q_\lambda=0.68$.
For $t \gsim 100$ days the evolution gets steeper and 
$\lambda_{10} = 0.81$ cm, and $q_\lambda=0.81$. The dispersion in the exponent is $\pm 0.03$. Similarly, the variation in $S'$
can for 10 -- 100 days   be fitted by 
\begin{equation}
S'(t) = S_{10} \left({t \over 10{\rm ~days}}\right)^{q_{\rm S}},
\end{equation}
where $S_{10} = 60.1$ mJy and $q_{\rm S}=2.43$,  while at later times ($\gsim
100$ days) the evolution is less steep  with $S_{10} = 1.67\EE2$ mJy
and $q_{\rm S}=2.00$.

Knowing  $\lambda_{\tau_{\rm ssa}=1}(t)$ and $S'(t)$, we can now
calculate the   magnetic field and the column density of relativistic
particles,
\begin{equation}
B(t) = 9.45\EE{-8} c(p)^2~ \left({R_s \over D}\right)^{4} S'(t)^{-2},
\label{eqbmags}
\end{equation}
where
\begin{equation}
c(p) = {(p+{7 \over 3}) \over (p+{10 \over 3})(p+1)} { \Gamma({p+8
\over 4})\Gamma({p+5 \over 4}) \Gamma({3p-1 \over 12})\Gamma({3p+7
\over 12})   \over  \Gamma({p+6 \over 4})\Gamma({p+7 \over
4})\Gamma({3p+2 \over 12})\Gamma({3p+10 \over 12})  }
\end{equation}
and $\Gamma$ is the gamma function. 
The  column density of relativistic particles is then given by
\begin{equation}
N_{\rm rel}(t) = {1.98\EE{17} \over (1.18\EE{-2})^{p}}~ { \Gamma({p+8
\over 4})  \over  (p+{10 \over 3})\Gamma({p+6 \over 4})\Gamma({3p+2
\over 12})\Gamma({3p+10 \over 12}) } B^{-(p+2)/2}~\lambda_{\tau_{\rm
ssa}=1}^{-(p+4)/2} ~{\rm cm^{-2}}.
\label{eqnrelq}
\end{equation}

As is seen from equations (\ref{eqbmags}) and (\ref{eqnrelq}), both
$B(t)$ and  $N_{\rm rel}(t)$ depend on $R_s(t)$,  which in turn is
determined by the ejecta structure. In this section we assume  
for simplicity a constant velocity of
 $V = 2.2\EE4 \kms$ (see section \ref{sec-velev}).

\begin{figure}  
\plotone{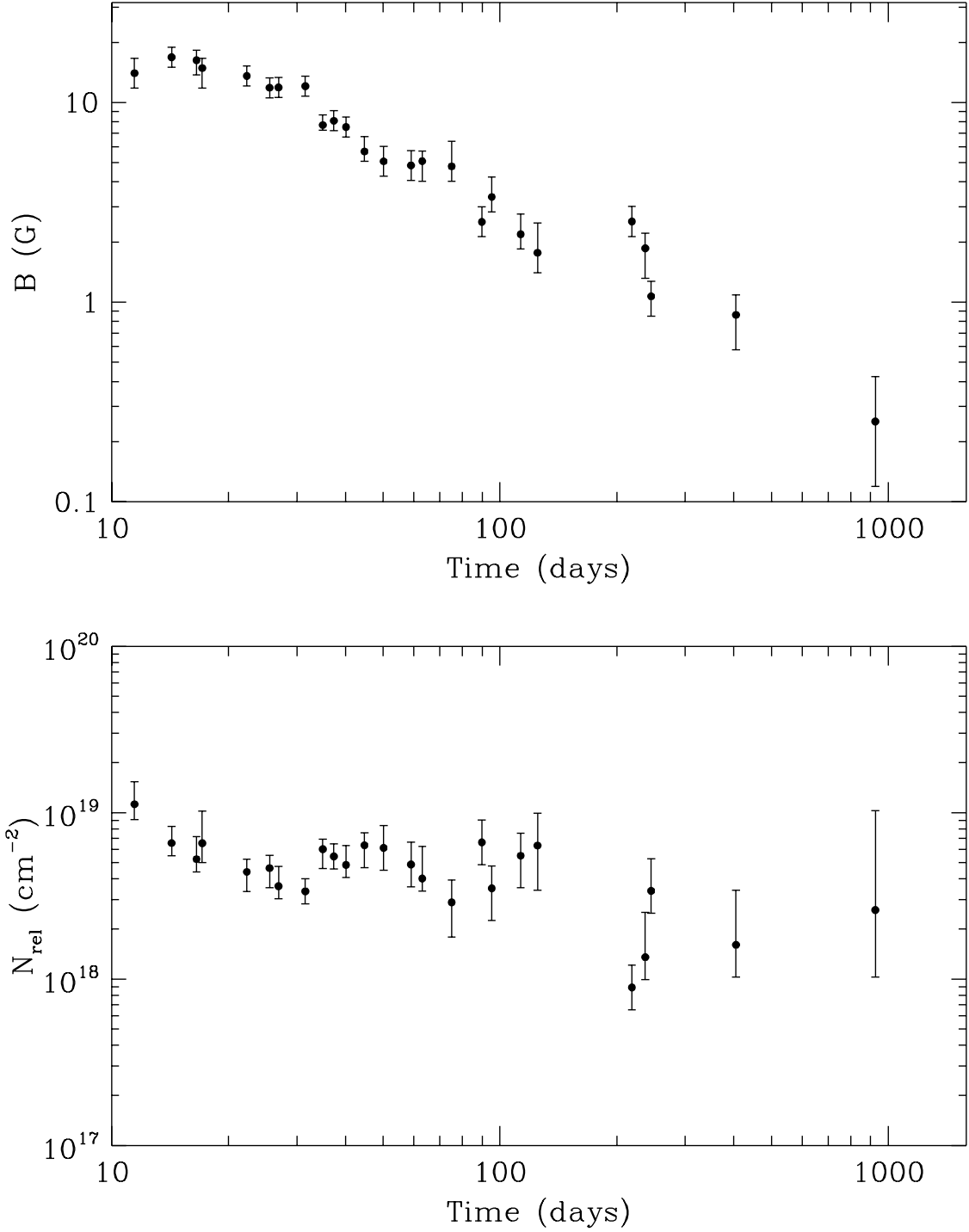}
\caption{Evolution of the magnetic field and column density of
relativistic electrons in the synchrotron self-absorption plus external 
free-free absorption model. } 
\label{fig5}
\end{figure}

The resulting magnetic field and column density are given in 
figure \ref{fig5}. 
During most of the observed interval the decrease in $B$ is
approximately a power law, and such a fit gives
\begin{equation}
B(t) = 25.5 ~ \left({t \over 10~{\rm days}}\right)^{-0.93 \pm
 0.08}
\left({V \over 2.2\EE4 \kms} \right)^{4} ~{\rm G}.
\label{eqeb}
\end{equation}

Within the errors, $N_{\rm rel}(t)$ is nearly constant, although the
scatter increases considerably for $t \gsim 100$ days.
The level of $N_{\rm rel}$ in figure \ref{fig5} should
be compared with that in the Razin model from section \ref{sec-onecomp}. Because of the higher
magnetic field,  and most importantly the smaller value of $p$, $N_{\rm rel}$ is roughly six orders of magnitude
smaller than for the Razin model.

With the magnetic field in equation (\ref{eqeb}) we can calculate the Razin wavelength 
from equation (\ref{eqh}) 
\begin{equation}
 \lambda_{\rm R0} = 3.5\EE2 ~\Mu5 \left({V \over 2.2\EE4 \kms}\right)^{2} 
\left({ t \over 10~ {\rm days}}\right) ~\rm cm.
\label{eqhq}
\end{equation}
Comparing equations (\ref{eqtaussa1})  and (\ref{eqhq}) we find that 
 the Razin effect in this model is unimportant.

Although the above results are of great interest, and will be discussed 
in detail below, there are several
deficiencies in the analysis. The most serious limitation is that
we have assumed a constant power law electron spectrum, 
i.e., neglected possible losses by radiation or collisions. We will
 therefore now,  
at the expense of a more complicated model, discuss a self-consistent 
analysis of the observations.    
The discussion in this section is, however, of great help in
order to understand the different processes involved, as well as a
guide to the values of $B$ and $N_{\rm rel}$ and their sensitivity to
different parameters.

\subsection{Self-consistent analysis}
\label{sec-selfc}
\subsubsection{Energy losses of the non-thermal electrons.}
\label{sec-invcomp}

For formulating a self-consistent analysis  it is necessary 
to discuss possible energy losses of the electrons which will affect 
the integrated electron
spectrum, and therefore also the emission from the plasma. 
To estimate these effects we assume in this subsection
that the magnetic field behaves like 
\begin{equation}
B(R_s) = B_{10} ~ ~ \left({t \over 10~{\rm days}}\right)^{-1}
\label{eqebmag}
\end{equation}
where $B_{10}$ parameterizes the strength of the field. From 
our earlier discussion and anticipating 
the results below, we expect
$B_{10} \approx 30$ G, to which  we scale our estimates. To
express our results as a function of time rather than radius we use in this section
$R(t) = 1.72\EE{15} (t/10~{\rm days})$ cm, corresponding to a constant 
velocity of $V = 2\EE4 \kms$.

We first estimate the typical Lorentz factor for the relativistic
electrons responsible for the radiation at wavelength $\lambda$ from
$\gamma \approx 85 ~(\lambda ~ B)^{-1/2}$, or
\begin{equation}
\gamma = 15.5~
\left({ t \over 10~ {\rm days}}\right)^{1/2} \left({B_{10} \over 30
 ~{\rm G}} \right)^{-1/2} \lambda^{-1/2}.
\label{eqgam} 
\end{equation}
The synchrotron lifetime for an electron to decrease its energy by a
factor two is 
\begin{eqnarray}
{t_{\rm synch} \over t}  = 
1.0~\gamma^{-1}
\left({ B_{10} \over 30 ~{\rm G}}\right)^{-2}
 ~ \left({ t \over 10~ {\rm days}}\right).
\label{eqw} 
\end{eqnarray}
 In terms of the observed wavelength this becomes
\begin{eqnarray}
{t_{\rm synch}  \over t} = 
6.4\EE{-2}~\lambda^{1/2}
\left({ B_{10} \over 30 ~{\rm G}}\right)^{-3/2}
   \left({ t \over 10~ {\rm days}}\right)^{1/2} 
\label{eqw2} 
\end{eqnarray}
It is obvious that synchrotron losses are important throughout most of
the observed period.

Because of the strong background radiation field from the supernova,
inverse
Compton losses may be important for the relativistic electrons. One can estimate
the timescale for this from
\begin{equation}
{t_{\rm Comp}  \over t}  = {3 \over 16 \pi} {m_e c^2 \over \gamma \sigma_{\rm T} J_{\rm Bol}}\approx 40~\gamma^{-1}~
\left({L_{\rm Bol} \over 10^{42} \ergs}\right)^{-1}
\left({V \over 2\EE4 \kms} \right)^{2}
\left({ t \over 10~ {\rm days}}\right),
\label{eqtcomp} 
\end{equation}
where $J_{\rm Bol}$ and $L_{\rm
Bol}$  are the bolometric mean intensity and 
luminosity, respectively. 

The bolometric luminosity has contributions from both the ejecta,
powered by the thermal energy from the passage of the shock and later
by radioactive decay, as well as  radiation produced by the
circumstellar interaction. The ejecta flux is emitted mainly in the
optical, and at early time also in the UV. 
We estimate the
ejecta luminosity 
from the bolometric luminosity calculated by Young, Baron \& Branch (1995),
 from observations by Richmond \etal (1994) and
Lewis \etal (1994). 
Unfortunately, the reddening is uncertain.
A range in $E_{\rm B-V}$ from 0.08 -- 0.32 has been estimated
by Richmond \etal
Unless otherwise stated we will use 
$E_{\rm B-V}=0.08$ in the following.

The luminosity in X-rays from the circumstellar and reverse shocks are
calculated in FLC96 and Fransson (1998). The circumstellar
shock is important only for $t \lsim 10$ days, and based on figure 8
in FLC96, scaled to $\Mu5 = 1$, we estimate $L_{\rm
cs} = 3.1\EE{41}~(t / 10~ {\rm days})^{-1.2} \ergs$. To include both
the in-going and outgoing radiation we multiply
the results in FLC96 by a factor two. 
In Fransson (1998) the evolution of the reverse shock luminosity is calculated 
for realistic density profiles, and it is  found that $L_{\rm rev} =
(1-2)\EE{41} \ergs$.   

The mean intensity, $J_i$, of the different luminosity components at the position of the 
circumstellar
shock is given by $J_i = W L_i / 4 \pi^2 R_i^2$, where $W = {1 \over 2} [1-(1 -
(R_{i}/R_{\rm s})^2)^{1/2}]$ is the dilution factor. The radius  $R_i$
stands for either the 
photospheric radius or the shock radius in the two cases. 
For $R_{ph} \ll R_{\rm s}$, $J \approx L_{\rm ej}/16 \pi^2 R_{\rm s}^2$. The radiation from
the shocks is more isotropic at the circumstellar shock.  Depending on
the thickness of the circumstellar shock and the location of the
acceleration zone, one expects that $R_{\rm s} \approx (1.0 -
1.3)~R_{\rm rev}$. 

\begin{figure}  
\plotone{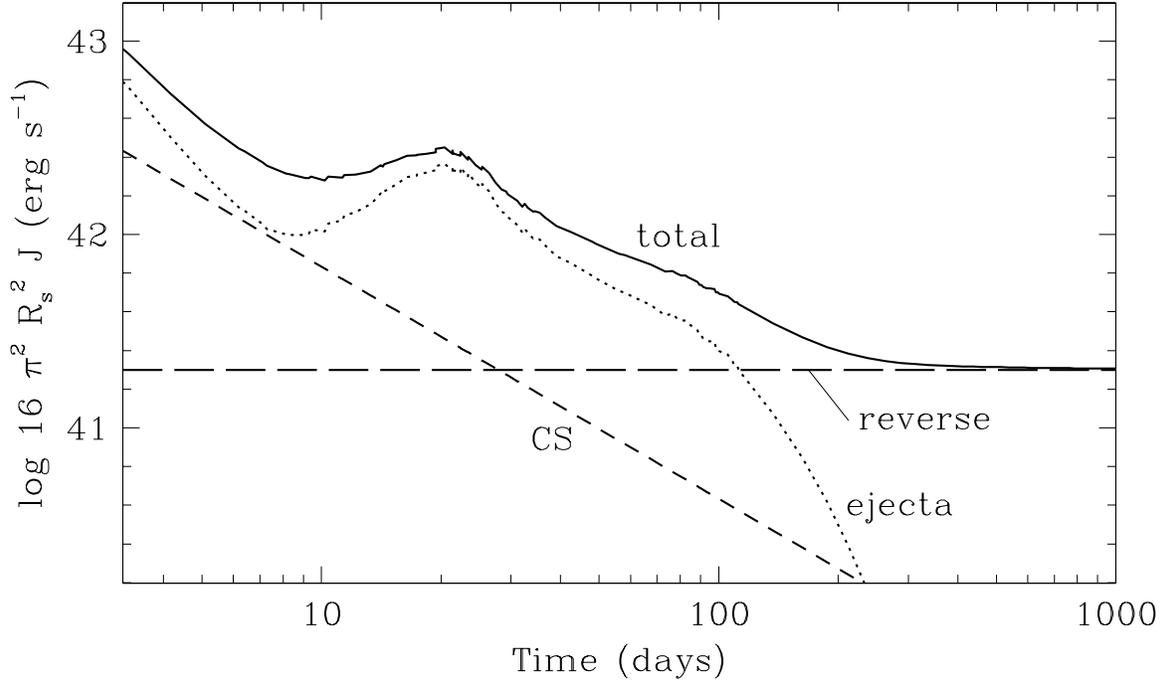}
\caption{Evolution of the mean intensity, $J$, at the shock, for
 $E_{\rm B-V} = 0.08$, $\Mdot = 5\EE{-5} \Msunyr$, and $u_w = 10 \kms$.
The quantity $16 \pi^2 R_{\rm s}^2 J$ is essentially 
the total luminosity at the shock. }
\label{fig6}
\end{figure}

In figure \ref{fig6} we show the resulting total
mean intensity multiplied by $16 \pi^2 R_{\rm s}^2$, 
together with the individual contributions from
the ejecta and the shock waves for $\Mu5 = 1$. In the case that $R_i
\ll R_s$ this  
essentially gives the total luminosity from component $i$. During the first $\sim 100$ days the radiation
 density
is dominated by the ejecta, with an increasing contribution from the reverse
shock, and can be approximated by $L_{\rm bol} \approx 4\EE{42} ~(t/10
{~\rm days})^{-0.9} \ergs$. 

 The Compton scattered optical photons from the ejecta and the thermalized X-rays from
the reverse shock emerge at
$h\nu \approx \gamma^2 ~h
\nu_0 \approx 0.1 - 100 {\rm ~keV}$. For our parameters the luminosity 
in this range is
 $\sim 2.5\EE{35} (L_{\rm synch} / 10^{37} \ergs) 
(L_{\rm bol} / 10^{42}~\ergs)
\ergs$. The
free-free flux from the shock (FLC96) therefore dominates the X-rays.
 One can also show that the ratio between first- and second-order Compton 
scatterings (Rees 1967) is small.
 
\begin{figure}  
\plotone{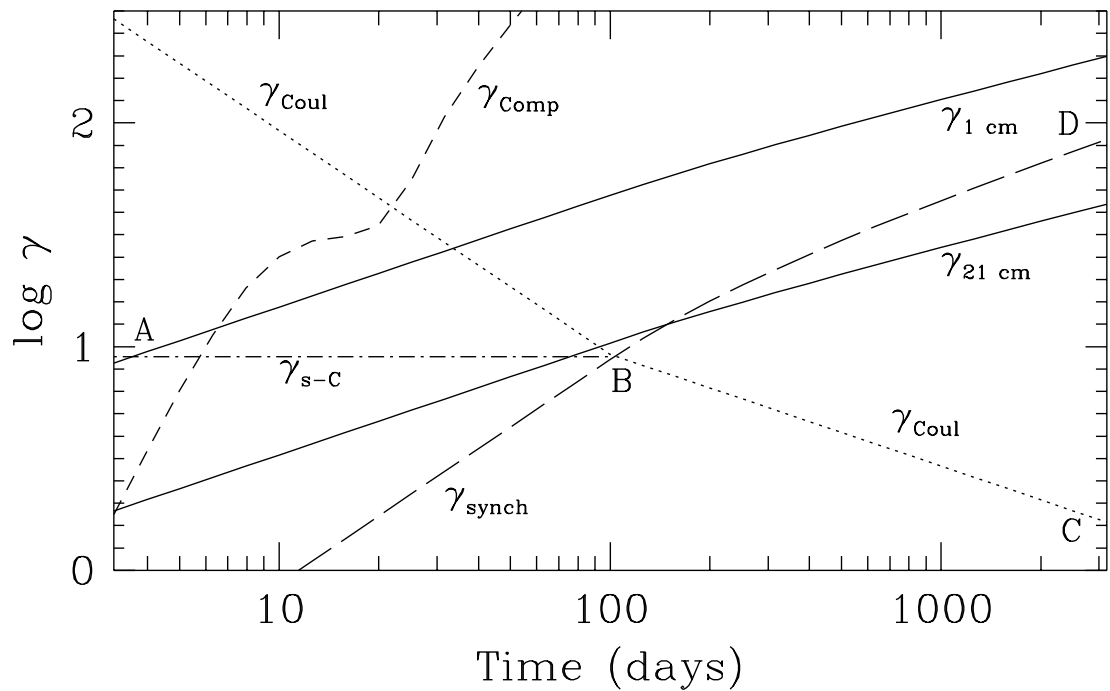}
\caption{Lorentz factor, $\gamma_{\rm synch}, \gamma_{\rm Comp}, \gamma_{\rm Coul}$, for which the
synchrotron, inverse Compton,  and Coulomb loss 
times, respectively, are equal to the expansion time. The
dot-dashed  line shows the region below which Coulomb losses are more
important than synchrotron losses. Below the 
line A -- B -- C Coulomb losses dominate,  and above the line A -- B -- D inverse synchrotron
losses dominate, while the region C -- B -- D is adiabatic.  The typical 
$\gamma$ responsible for the
emission at 1 cm and 21 cm are shown as solid lines.  }
\label{fig7}
\end{figure}
\begin{figure}  
\plotone{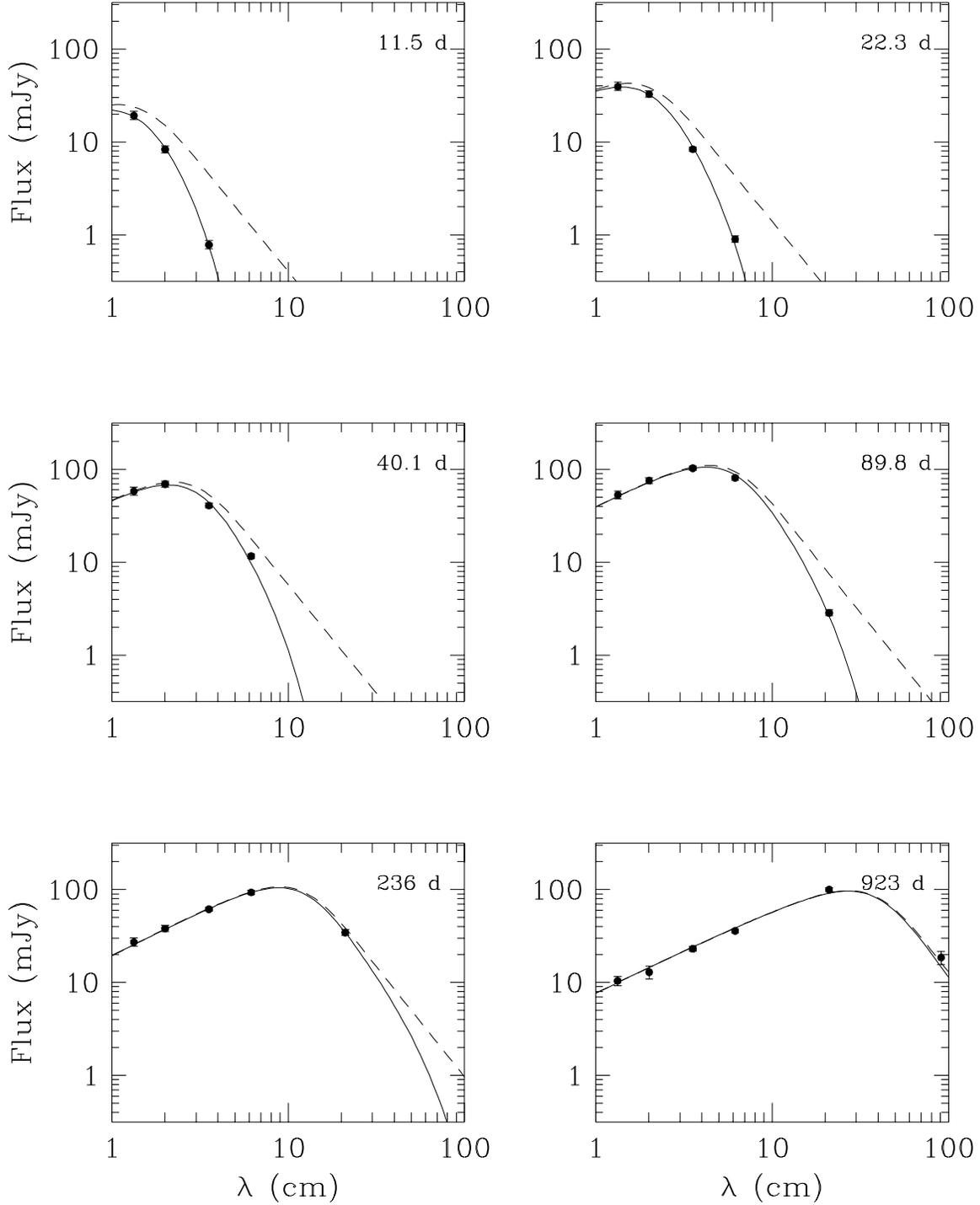}
\caption{Fits to the VLA spectra at six different epochs, including all
suppression processes, as 
well as a self-consistent calculation of the electron spectrum.   The
dashed line shows the intrinsic spectrum from the 
shock, while the solid line also include external free-free absorption.}
\label{fig8}
\end{figure}

Knowing both the magnetic field and the radiation density, we can
calculate the critical value of the  Lorentz factors, $\gamma_{\rm synch}$ and
$\gamma_{\rm Comp}$, for which the synchrotron and Compton timescales are equal to
the expansion timescale. In figure
\ref{fig7} this is shown as the solid line. We also show the approximate value of $\gamma$
 for the electrons
responsible for the emission at  1 cm and 21 cm, $\gamma_{1 ~{\rm
cm}}$ and $\gamma_{21 ~{\rm cm}}$,  respectively, as given by equation (\ref{eqgam}).
Clearly,  synchrotron losses are more important than inverse Compton
losses at all times. 

Coulomb losses can also be
important for the relativistic electrons. For these electrons the  
ratio of the Coulomb timescale to the expansion timescale is given by 
\begin{eqnarray}
 {t_{\rm Coul} \over t} = 8.9\EE{-3}~\gamma~\Mu5^{-1}
  ~ \left({V \over 2\EE4 \kms}
\right)^{2}  \left({ t \over 10~ {\rm days}}\right).
\label{eqaw} 
\end{eqnarray}
For $\gamma \lsim \gamma_{\rm Coul} = 1.1\EE2~(t /10~{\rm days})^{-1}$
Coulomb losses are important. We should also compare
$t_{\rm Coul}$ with the synchrotron timescale, and we find that for 
\begin{equation}
\gamma  \lsim \gamma_{\rm s-C} = 8.62 ~\Mu5^{1/2}
 ~\left({B_{10} \over
30 \rm ~G}\right)^{-1}
 \left({V \over 2\EE4 \kms}
\right)^{-1}
\label{eqay} 
\end{equation}
Coulomb losses dominate synchrotron losses. In figure \ref{fig7} we show $\gamma_{\rm Coul}$
and $\gamma_{\rm s-C}$ for $\Mu5 = 1$. We note that $\gamma_{\rm s-C}$ is
independent of time, and that especially at early times Coulomb cooling
is important for the longer wavelengths.

Synchrotron and inverse Compton losses steepen
the electron spectrum index by one (e.g., Pacholczyk 1970, and section \ref{sec-lum}
below). 
Coulomb losses have the opposite
effect to synchrotron losses in that the electron spectrum index flattens
by one power, compared to the injected spectrum.
Therefore, with the parameters in figure \ref{fig7} we expect  that before $\sim 100$ days (i.e., when $\gamma_{\rm synch} \sim \gamma_{\rm Coul}$) 
the electron spectrum will be flattened by one power for energies below $\gamma_{\rm s-C}$ 
and steepened by one power above $\gamma_{\rm s-C}$. 
In reality, 
both Coulomb and synchrotron compete in the regions responsible for the observed
emission, 
and the spectral index is intermediate between these values. 
Later than $\sim 100$ days the electron 
spectrum 
will have three regions, one between $\gamma_{\rm Coul} < \gamma < \gamma_{\rm synch}$,
where the spectrum is equal to the original input spectrum, while below and above this region 
the spectrum is flattened and steepened by one power in energy,
respectively. We return to this issue in section \ref{sec-lum}.

\subsubsection{Calculation of the electron spectrum}
\label{sec-elspec}

For the acceleration of the relativistic electrons, once they have been
injected from the thermal pool, there are at least two possibilities. First-order Fermi acceleration across the circumstellar
shock is an obvious candidate, but acceleration in the turbulent region
close to the contact discontinuity is also a possibility 
(Chevalier, Blondin, \&
Emmering 1992). The timescale of acceleration depends on the spatial
diffusion coefficient, and therefore on the mean free paths. Only rough
estimates of the timescale based on modeling of the radio emission from SN 1987A are available
(Ball \& Kirk 1992; Duffy, Ball, \& Kirk 1995). 
From the expression given by  Ball \& Kirk (1992) for
the variation of flux with time (their equation 8) we find that the rise is 
likely to be very rapid, $t_{\rm acc} \ll 10$ days.
Here, we therefore 
assume that the electrons are accelerated almost instantaneously (i.e.
$t_{\rm acc} \ll t_{\rm Comp}, t$), with a spectrum $ \propto
\gamma^{-p_{\rm i}}$. 

In calculating the electron distribution we use the equation
\begin{equation}
{\partial N(\gamma) \over \partial t} = {\partial \over \partial \gamma} 
[ {\gamma \over t_{\rm loss}} N(\gamma) ] + {\partial \over \partial
\gamma}[E(\gamma) \gamma^2 {\partial \over 
\partial \gamma}({N(\gamma) \over \gamma^2}) ] -
{N(\gamma) \over t_{\rm esc}} + 
{(p_{\rm i}-1) V n_{\rm rel} \over 4 \gamma_{\rm min}} 
\left( {\gamma_{\rm min} \over \gamma} \right)^{p_{\rm i}}.
\label{eqzqqa} 
\end{equation}
The first term on the right hand side takes all energy losses into
account, with a total energy loss timescale  given by 
\begin{equation}
t_{\rm loss} = \left({1 \over t_{\rm synch}} + 
{1 \over t_{\rm Comp}} + {1 \over t_{\rm Coul}} + {1 \over t}\right)^{-1}.
\label{eqzqqb} 
\end{equation}
The first three terms on the right hand side are already defined,
while the last term takes adiabatic losses into account. 
In equation (\ref{eqzqqa}) the second term, where $E(\gamma)$ is given
by
\begin{equation}
E(\gamma) = {1 \over 2 m_{\rm e}^2 c^2} \int j(\nu, \gamma) 
{J_{\nu} \over \nu^2} d\nu,
\label{eqzqqc} 
\end{equation}
accounts for the heating of
the non-thermal electrons by the radiation. In this $j(\nu, \gamma) $ is the
single electron emissivity.
The third term describes the escape of particles, either as a result
of  advection towards the contact discontinuity,
and subsequent energy degradation by collisions in the high density gas, 
or as a result of actual escape. Finally, the last term gives the number of injected electrons with
energy $\gamma m_{\rm e} c^2$ per second and unit area, 
which we assume to be instantaneously
accelerated at the shock. A similar equation was first
used by McCray (1969) to discuss the thermalization of the electron
distribution, when synchrotron self-absorption is important, while
the form given here is basically that used by de Kool, Begelman, \&
Sikora (1989). The equation assumes that the electrons are 
relativistic, but given our ignorance of the details of the injection
process, which 
certainly influences the distribution for $\gamma \sim 1$, this is
sufficient. Equation (\ref{eqzqqa}) is solved by the semi-implicit
method of Chang \& Cooper (1970).

If we assume that the self-absorption effects are small and that the 
escape term can be ignored to a reasonable approximation, as is confirmed by the detailed calculations in section 
\ref{sec-lum}, we can write 
down the stationary solution to equation (\ref{eqzqqa}) as 
\begin{equation}
N(\gamma)  = {n_{\rm rel} V t \over 4 \gamma_{\rm min}}   
\left({\gamma \over \gamma_{\rm min}}\right) ^{-p_{\rm i}}
(1 + {t \over t_{\rm Coul}} + 
{t \over t_{\rm Comp}} + {t \over t_{\rm synch}})^{-1}.
\label{eqnrel}
\end{equation}
With this electron spectrum we can calculate the synchrotron emissivity and opacity 
from the general expressions 
in equations (\ref{eql})  and (\ref{eqla}),
as well as the source function, and finally the synchrotron spectrum as function 
of time. 

Because $V(t)$, $L_{\rm bol}$ and  $\Mdot$ are determined from other
observations (section \ref{sec-velev}), we have
for each spectrum two parameters,
$n_{\rm rel} $ and $B$. We have to assume a specific {\it
input} spectrum for the electrons (i.e., $p_{\rm i}$), 
which we take to be constant with time.

\subsubsection{Velocity evolution}
\label{sec-velev}
The magnetic field and electron density depend on the velocity evolution of
 the shock, i.e., on $R_s$  (e.g., equation \ref{eqbmags}). 
The velocity of the shock region is in turn sensitive to  the density structure of the ejecta. 
If the  ejecta density is
approximated by a power
law with an exponent $n$, the shock radius varies as $R_{\rm s}
\propto t^m$, where, for an $n_{\rm e} \propto r^{-2}$ wind, $m = {(n-3)/(n-2)}$ (Chevalier
1982a). A flat density profile, $n \lsim 7$, causes the shock to slow down,
while a steep profile,  $n \gsim 20$,  gives a shock of fairly constant
velocity. 
Unfortunately, the
density structure of the ejecta is uncertain, depending on an equally uncertain
progenitor structure. Optical line profile observations of SN
1993J
provide some constraint on $V(t)$.
This, however, refers to the ejecta gas, and not directly to the radio
 emitting plasma, which is likely to be $20 - 30 ~\%$ further out.  
Also, if there is a change in the density gradient, one can expect a
 lag of the adjustment to the new gradient.

The most direct information about the size of the radio emitting 
region comes from VLBI observations. During the first 92 days  Bartel \etal (1994) found 
roughly undecelerated expansion, $m=0.96 \pm 0.07$, with a velocity $V =19000 \pm 3000 \kms$. 
Most of the uncertainty in $V$  comes from the M81 distance errors,
$D= 3.63 \pm 0.34$ Mpc  
(Freedman \etal 1994), which do, however, not 
enter in $m$.  At later times, $182 - 1304 $ 
days,  Marcaide \etal (1997) find strong evidence for deceleration, with $m\approx0.86 \pm 0.02$.
It is quite possible that the VLBI velocities above are underestimated. 
FLC96, as well as Bartel \etal, argue 
that the $\Ha$ line indicated an ejecta velocity of 18000 -- 19000
$\kms$ at $10 - 30$ days. The shock velocity should be a factor of
$1.2 - 1.3$ larger than this, or $21600 - 24700 \kms$. Furthermore, 
the VLBI measurements have a fairly small dynamic range;  
faint high velocity material may therefore go undetected. We also point out that the radius measured by Marcaide \etal only refers to the 
50 \% contour level, which explicitly excludes such faint, high velocity
material. Guided by this discussion we will in the following 
parameterize the velocity by
$$
V(t) =  \cases{
V_0 &~$t < t_{\rm p}$ \cr
V_0~\left({t \over t_{\rm p}}\right)^{(m-1)} &~$t_{\rm p} \le t$,
}
\label{eqvela} 
$$
with $V_0=22000 \kms$. Because we take $V =$ constant for $t < t_{\rm p}$ our value of
$m$ will be less than the value of $m=0.86$ from Marcaide \etal With
$t_{\rm p} = 100$ days our effective value of $m$ up to day
1304 is $m=0.74$, corresponding to $n \approx 6$. Using  $t_{\rm p}=200$ days
would not  change our results significantly. 

\section{RESULTS}
\label{sec-disc}
With all parameters fixed, except $B$, $n_{\rm rel}$ and $p_{\rm i}$, 
we can now
repeat the modeling in section \ref{sec-ssaff} for each date, thereby
determining the unknown parameters from a $\chi^2$ fit. However, in 
contrast to section \ref{sec-ssaff} we take the different cooling 
processes into account. The resulting spectra at six epochs, from 11.5 -- 923 days, are 
shown in figure \ref{fig8}. Also shown is the spectrum when 
free-free absorption is neglected. It is seen that when the latter process 
is included the fit is essentially perfect, as is the case also for the other 
dates. We now discuss
the implications for the physical parameters resulting from this procedure.


\subsection{Free-free absorption}
\label{sec-freefree}
The analysis in section \ref{sec-ssaff} shows that the
best fits for the free-free absorption  are obtained for $\delta=1.0$ and 
$T_{15} = 2.75\EE6$ K and an asymptotic temperature of $2\EE5$ K. 

This, and the evolution of temperature derived from the
observations, can be compared to the calculations in FLC96. 
At 10 days FLC96 (their figure 11) find that at a radius of $1.5
\times R_s$, $T_e \approx 6.3\EE5$ K, decreasing to 
$3.0\EE5$ K at 20 days and then staying at a constant value of $2.5 \EE5$ K
 up to day 100. 

\begin{figure}  
\plotone{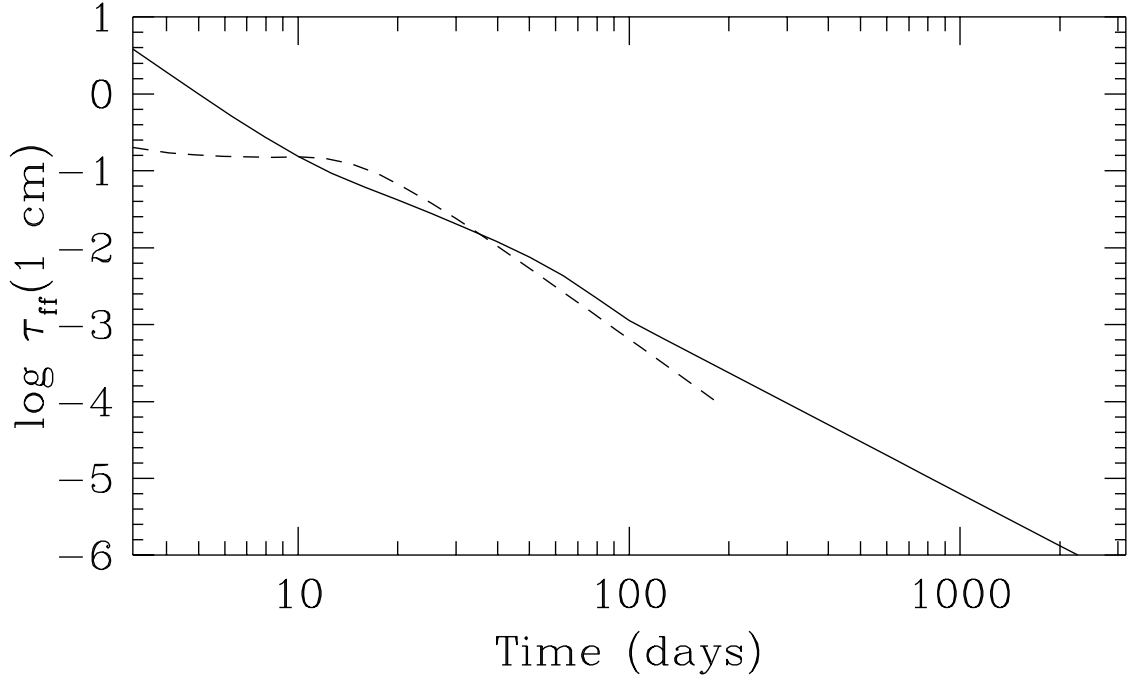}
\caption{The free-free optical depth at 1 cm determined from the fits in this
paper, compared to the optical depth from the model calculations of the
circumstellar temperature in FLC96 (dashed line).}
\label{fig9}
\end{figure}

As we stressed in section \ref{sec-ssaff}, the most relevant quantity to 
compare with the observations is 
not the temperature, but the the optical depth. 
A further complication is that the temperature in FLC96 not only varies 
with radius, but, contrary to our
model assumptions, it
also shows some variation with time at a given radius. For a real
comparison one should perform 
the radial integration of the optical depth at each time (i.e., $\tau_{\rm ff} 
\propto
 ~\int_{R_s}^\infty T_{\rm e}(r,t)^{-1.5} n_{\rm e}(r)^2 dr$).
In figure 
\ref{fig9} we 
show the free-free optical depth, averaged over the disk, for the $\rho
\propto r^{-2}$  model in FLC96 along with that determined by the
 observations. In
general the  two results agree to better than a factor of two. 
While the 
difference is significant, we feel that it is remarkable that they agree 
as well as they do, and that this agreement gives us increased confidence 
in the determination of the circumstellar temperature from the model 
calculations not only in the case of SN 1993J, but also for other objects 
like SN 1979C and SN 1980K (Lundqvist \& Fransson 1988). 
An approximate fit to our results in figure \ref{fig9} is given by  
\begin{equation}
\bar{\tau}_{\rm ff}(t) = 0.12 ~\lambda^{2.1}~
\left({t \over 10~{\rm days}}\right)^{-2.1}.
\label{eqt2} 
\end{equation}
The fact that we get a power law significantly flatter than $t^{-3}$ 
even in the constant temperature region at $t \gsim 100$ days is 
due to the compensating effect of the 
decreasing velocity, which gives an approximate power law dependence over 
the whole period.

Although in
broad agreement with the optical depth required by our modeling, there
are considerable uncertainties in the calculation of the temperature
of the circumstellar medium, as is discussed in FLC96. Most important
of these are the degree of equipartition of electrons and ions behind
the circumstellar shock, and the evolution of the shock velocity
shortly after break-out. Both affect the temperature of the shocked
electrons, as well as the electron scattering optical depth. These two
quantities determine the UV and soft X-ray flux from the shock through
Comptonization of the soft photospheric photons. The uncertainty in
the shocked electron temperature is therefore reflected in the
absolute value and the radial dependence of the temperature of the circumstellar medium. 
The radio observations therefore provide
a new constraint on the physics and radiation mechanisms of the shock. 

\subsection{Magnetic field}
\label{sec-magn}

\begin{figure}  
\plotone{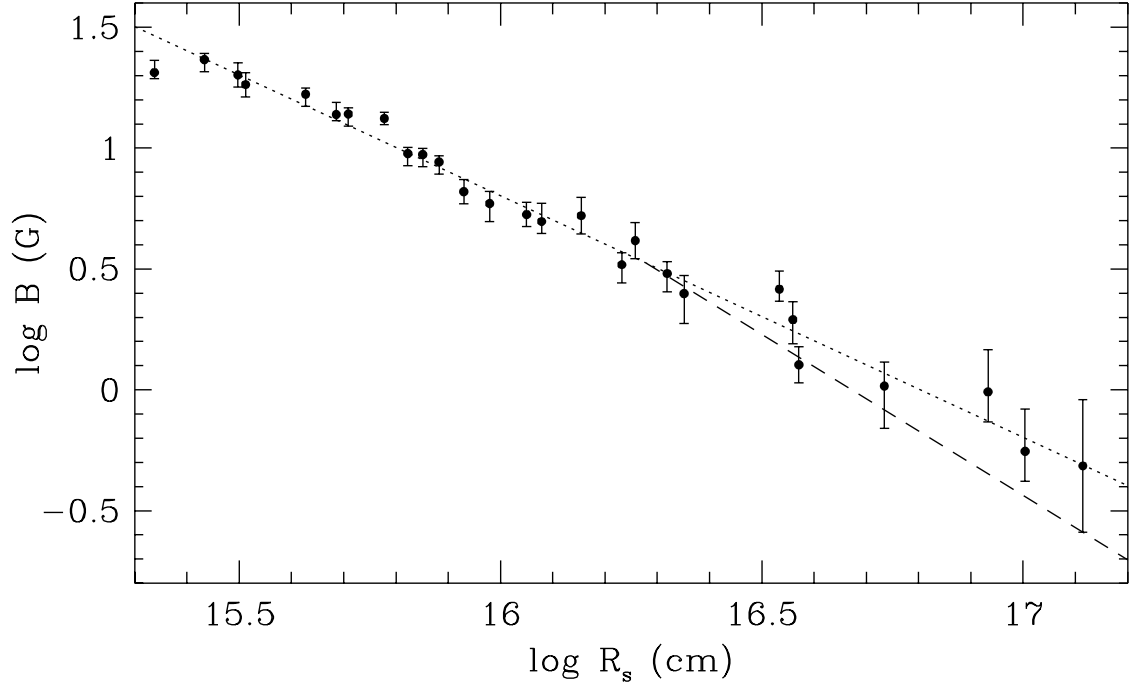}
\caption{Evolution of the magnetic field as a function of shock
radius. The dashed  line shows the expected evolution if $B^2 \propto
\rho_{\rm wind} V_{\rm s}^{2}$, while the dotted line
shows the evolution of an $B \propto R_{\rm s}^{-1}$ field.}
\label{fig10}
\end{figure}

In figure \ref{fig10} we show the resulting magnetic field as a function of the shock radius. 
The errors in $B$ and  $n_{\rm rel}$ are 
determined from the $1 ~\sigma$ contours of constant $\chi^2$ in the  $B$ -- $n_{\rm rel}$ 
plane.
The scatter in $B(t)$ at late times is  related to the scatter in $S'(t)$, 
which enters 
quadratically in $B(t)$. 

A least squares fit to the data shows that 
\begin{equation}
B(t) = B_{15} ~ \left({R_s \over 10^{15}~{\rm cm}}\right)^{-\beta}
\left({V_{0} \over 2.2\EE4 ~\kms}\right)^{4} .
\label{eqebma}
\end{equation}
with $B_{15} = 61.1 \pm 4.7$ G and $\beta = 0.984 \pm 0.028$.
As seen by the dotted line in figure \ref{fig10}, it is
certainly compatible with a  simple $B \propto R_s^{-1}$ law. 
A best  fit with this dependence is given by $B_{15} = 63.5 \pm 4.8$ G
($\beta \equiv 1.0$).  
Note that the errors do not include systematic errors. In particular,
the results are, as shown 
by the expressions above, sensitive to the velocity evolution, as well
as its absolute value.  

A $B \propto R_s^{-1}$ dependence is  close to that expected if the magnetic field 
behind the shock is amplified 
by a constant factor from the circumstellar magnetic field of a
rotating progenitor. 
If the wind has sufficient angular momentum, the magnetic field should have the 
topology of a 'Parker spiral' at large
radii (Parker 1958; Weber \& Davis 1967). In 
this case one expects $B \propto r^{-1}$, while a radially expanding wind has $B \propto 
r^{-2}$.

 The magnetic fields of the
circumstellar media of late type supergiants are uncertain. Based on polarization
observations of OH masers in supergiants, Cohen
\etal (1987) and Nedoluha \& Bowers (1992)  estimate  that at a radius of
$\sim 10^{16}$ cm the magnetic field is $\sim 1 - 2 {~\rm mG}$, though 
the uncertainty
in this number is large. It is unlikely that the magnetic field in the wind is higher than that
corresponding to equipartition between the magnetic field and the kinetic energy of
the wind. This means that ${B^2 \over 8 \pi} \lsim \rho~u_{\rm w}^2 /2 $, giving
\begin{equation}
B \lsim {(\Mdot~u_{\rm w})^{1/2} \over r} = 
  2.5~\left({\Mdot\over 10^{-5}
\Msunyr}\right)^{1/2}
\left({ u_{\rm w} \over 10 \kms}\right)^{1/2} \left({ r \over 10^{16}~
{\rm cm}}\right)^{-1} ~{\rm mG}
\label{eqt} 
\end{equation}
Likely locations for the
electron acceleration are at the
position of the circumstellar shock, or, alternatively, close to the contact
discontinuity between the circumstellar swept-up gas and the shocked
ejecta gas. The latter region is Rayleigh-Taylor unstable, and the associated turbulence
may help amplifying the magnetic field (Chevalier, Blondin \&
Emmering 1992; Jun \& Norman 1996). 

At 10 days, corresponding to a radius $\sim 1.9\EE{15}$
cm, we find that the
magnetic field in the emitting region is $\sim 34$ G and the . 
Using the above estimate of the circumstellar magnetic
field and a shock compression by a factor four, this post-shock  magnetic 
field  would be
 $B \approx
(2.4-4.8)\EE{-2}$ G. This is a factor $\sim 10^3$ less than that inferred
from the observations,  and therefore strongly argues for 
magnetic field  amplification behind the
shock. Although this conclusion rests on the very uncertain estimate of the
circumstellar magnetic fields of the progenitor system, a simple shock compression of
the field can probably  be excluded, and gives support to the results found in the
simulations by Jun \& Norman (1996).

If the scaling of the magnetic field is based on the thermal energy density one expects $B^2/8 \pi \propto \rho V^2$, implying 
$B \propto V/R_s \propto t^{-1} \propto R_s^{-1/m} $ (e.g., Chevalier
1996). This is shown as the dashed line in figure
\ref{fig10}. Although 
there is formally better agreement with the $B \propto R_s^{-1}$ curve
(dotted line), the errors in the 
data in 
figure \ref{fig10} are large enough that we find it difficult to
separate these two cases. 
In the following discussion we will assume that the magnetic field
follows  
\begin{equation}
B(t) = B_{10} ~ \left({t \over 10~{\rm days}}\right)^{-1}
\left({V_0 \over 2.2\EE4 ~\kms}\right)^{4} ,
\label{eqebmat}
\end{equation}
with $B_{10} = 34$ G. This choice is of no importance for the subsequent analysis, and
is only argued  from the point of view of the possible importance of
equipartition. 

The ratio of the magnetic energy density to the thermal energy density
behind the shock is 
\begin{equation}
{u_{\rm B} \over u_{\rm therm}} \approx 0.14 
\left({B_{10} \over 34~{\rm G}}\right)^2~ 
\dot{\cal{M}}^{-1} 
\label{eqdrat}
\end{equation}
 independent of time.
It is interesting that the field is of the same order as that given by
equipartition. The difference from equipartition is, however, significant.

\subsection{Injection rate of non-thermal electrons} 
\label{sec-nonth}

\begin{figure}  
\plotone{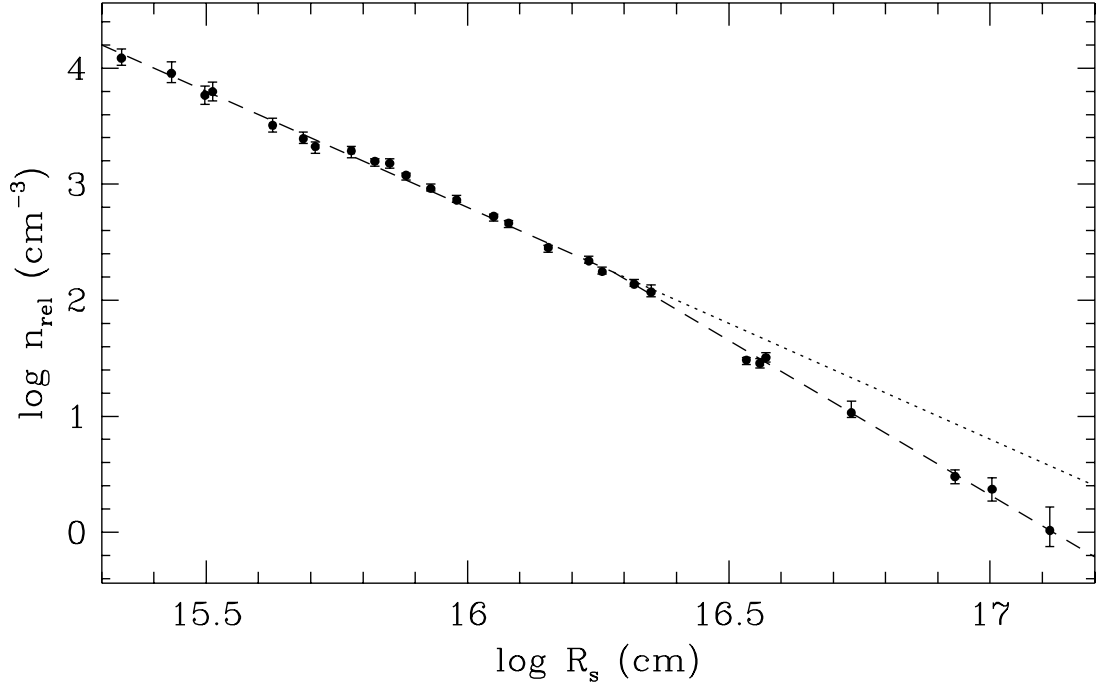}
\caption{The number of relativistic particles as a function of the shock
radius. The dashed  line shows the expected evolution if $n_{\rm rel} \propto
\rho_{\rm wind} V_{\rm s}^{2}$, while the dotted line
shows the evolution if $n_{\rm rel} \propto \rho_{\rm wind}$.}
\label{fig11}
\end{figure}

As we discuss in more detail in the next section, the power law index of the injected
electron spectrum is close to $p_{\rm i} = 2.1$, and constant in time. We therefore
use this value in the discussion below.  In figure 
\ref {fig11} we show the density of the injected non-thermal electrons,  given by $n_{\rm rel}$, as a
function of  shock radius. 
The value of $n_{\rm rel}$ is determined mainly by the optically thin 
flux, and therefore, from the arguments in section \ref{sec-lum},  can be shown to depend on velocity as $n_{\rm rel} 
 \propto V^{3-2 p_{\rm i}} \propto V^{-1.2}$. 
A least squares fit to the data in figure \ref {fig11} 
for the first 100 days  is given  by
\begin{equation}
n_{\rm rel}= n_{\rm rel~15} ~ \gamma_{\rm min}^{-1.1} \left({R_s \over 10^{15}~ 
{\rm cm}}\right)^{-\eta} 
\left({V_0 \over 2.2\EE4 ~\kms}\right)^{-1.2},
\label{eqdrelr}
\end{equation}
where  $n_{\rm rel~15} = (6.1 \pm 0.7)\EE4~\ccm$ and $\eta = 1.98
\pm 0.04$.  There is after 100 days a prominent steepening of the slope, 
and one finds that $n_{\rm rel~15} = (4.0 \pm 0.9)\EE5~\ccm$ and $\eta 
= 2.64 \pm 0.05$.
A fit based on $\eta = 2$ during the first 100 days gives  $n_{\rm
rel~15} = (6.4 \pm 0.8)\EE4~\ccm$.

Chevalier
(1996) has discussed  scalings for the number density of relativistic 
particles based on either a constant fraction of
the thermal particle density, 
$n_{\rm rel} \propto \rho_{\rm wind} \propto R^{-2}$, or  a constant fraction of
the thermal energy density, 
$n_{\rm rel} \propto \rho_{\rm wind}~V^2 \propto R^{-2} ~V^2\propto t^{-2}$. 
Here $\rho_{\rm wind}$ is the wind density.
These scalings have  little physical motivation, and can only be
justified by  observations. An
argument in favor of these  scalings is that 
similar scalings have reproduced the light curves of other
supernovae dominated by free-free absorption.
We can now test these scalings directly against the observations in the same 
way as the magnetic field. 
 Taking the two
alternatives for the relativistic particle injection efficiencies into account, 
we write 
\begin{equation}
n_{\rm rel} = f_{\rm rel}~ n_{\rm e} \left({V \over V_0}\right)^{2
\epsilon},
\label{eqzqq} 
\end{equation}
where $\epsilon= 0$  for $n_{\rm rel} \propto \rho_{\rm wind}$
 and $\epsilon = 1$ for $n_{\rm rel} \propto \rho_{\rm wind}~V^2$. The
thermal electron density, $n_{\rm e}$, behind the shock is given by
 equation (\ref{eqac}). 

As long as the velocity is constant both 
prescriptions give the same result, $n_{\rm rel} \propto R^{-2}$. 
When the velocity decreases after 100 days, $\rho_{\rm wind}~V^2 \propto 
t^{-2} \propto R_{\rm s}^{-2/m} \propto R_{\rm s}^{-2.66}$ for $m=0.74$. 
This is shown as the dashed line in figure \ref {fig11}.
As is obvious from both the figure and the fit above, $n_{\rm rel} \propto 
\rho_{\rm wind}~V^2$ reproduces the observations remarkably well,
while the $n_{\rm rel} \propto \rho_{\rm wind}$ model is incompatible
with the observed evolution. 
 This important result shows that {\it a fixed fraction, $ f_{\rm rel}$, of the thermal particle 
energy density is accelerated to relativistic energies}. 

The best fit value of $f_{\rm rel}$ found by this analysis and the
modeling in next section is $f_{\rm rel}=1.85\EE{-4}$. 
The energy density in relativistic electrons immediately behind the shock 
is 
\begin{equation}
u_{\rm rel} = f_{\rm rel} n_{\rm e} m_{\rm e} c^2 
{(\gamma_{\rm min}^{-p_{\rm i}+2}-\gamma_{\rm max}^{-p_{\rm i}+2}) 
\over (p_{\rm i}-2)}
\equiv
f_{\rm rel} n_{\rm e} m_{\rm e} c^2 Q(\gamma_{\rm min},\gamma_{\rm
max},p_{\rm i}). 
\label{eqeff} 
\end{equation}
The ratio of the non-thermal electron
density and the thermal
total energy  density behind the shock is
$u_{\rm rel}/ u_{\rm therm} \approx 5.9\EE{-5}~ Q$. For $p_{\rm i} = 2.1$, 
$\gamma_{\rm min} = 1$ and $\gamma_{\rm max} =
\infty$, $Q =8.7$, while 
$\gamma_{\rm min} = 5$ and $\gamma_{\rm max} = 15$,  corresponding 
to the directly
inferred range at 10 days, gives $Q = 0.9$. In either case the 
non-thermal {\it
electron} density is much smaller than the thermal electron density, 
in contrast to the
Razin case. There may, however, be a substantial
energy density of  non-thermal {\it ions} accelerated by the shock. 
For an equal number
density of ions one then gets $u_{\rm ion} / u_{\rm therm} \approx 0.11 ~Q$. It is therefore
conceivable that, to within an order of magnitude, rough equipartition
between the non-thermal particles, the magnetic
field and the thermal energy is present behind the shock. This
conclusion rests, however, totally on the ad hoc assumption
of equal efficiency for accelerating electrons and 
ions.

\section{RADIO LIGHT CURVES FOR SN 1993J}
\label{sec-lum}

The most direct way to compare the model with the observations is through the 
radio light curves at
the different wavelengths. By calculating the time evolution
we can also include other effects that were not possible for the
analysis above. It therefore serves as a check on the assumptions 
of this analysis. 
In contrast 
to the analysis in sections \ref{sec-magn} and \ref{sec-nonth}, we now make an assumption about the the behavior of the 
magnetic field and injection efficiency. 
For the injection rate we use equation (\ref{eqzqq}), with $\epsilon =
0$ or $\epsilon = 1$.
The high ratio of $u_{\rm B}/ u_{\rm therm}$ (equation \ref{eqdrat}) may 
be taken as an argument for a scaling $u_{\rm B} \propto u_{\rm
therm}$, or $B \propto t^{-1}$, rather than $B \propto R_{\rm s}^{-1}$, and 
the calculations below therefore assume  $B \propto t^{-1}$. As
we   show, the results are essentially independent of this
assumption. 
For the radiative transfer we use equations (\ref{eqssa}),
(\ref{eql}),  and (\ref{eqla}). In contrast to section \ref{sec-disc}, we now 
use the full form of equation (\ref{eqzqqa}), including self-absorption 
effects on the electron distribution, as well as advection and the
explicit  time-dependence. 

The evolution  of the velocity of the shock (through $B$, $n_{\rm rel}$ and $R$) is
 important for
the  number of relativistic electrons, and we use the 
 parameterization in section \ref{sec-velev}.  
In the analysis we include all VLA observations, as well as the 2 cm 
observations by Pooley
 \& Green (1993). In particular, this includes 
several observations from 5 - 10 days, not used in the analyses in 
sections \ref{sec-ssaff} and \ref{sec-disc}.

\begin{figure}  
\plotone{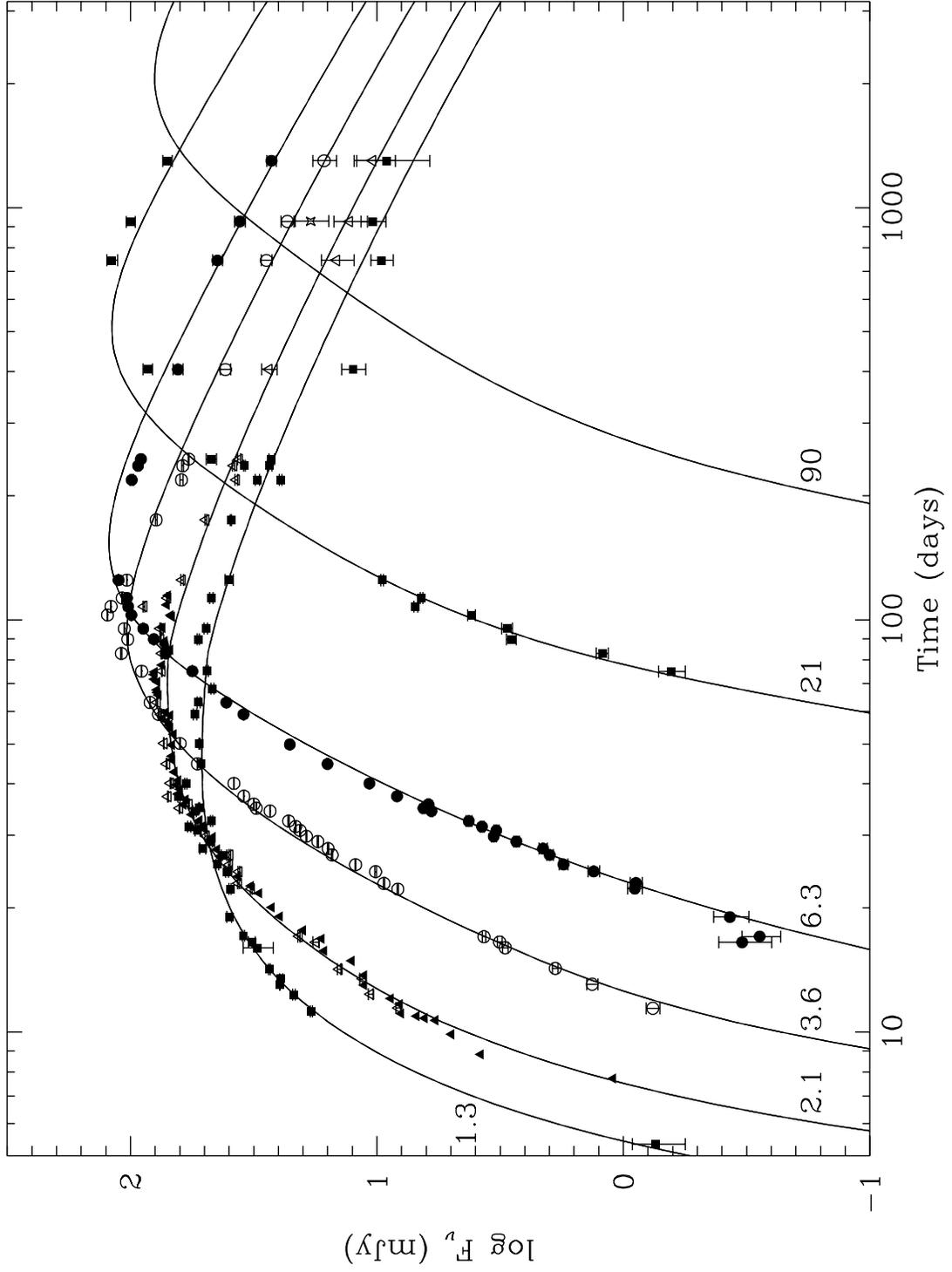}
\caption{Calculated and observed radio light curves 
for the model with $p_{\rm i} = 2.1$,
$\epsilon = 1$, $f_{\rm rel} = 1.85\EE{-4}$ and $n=6.5$, discussed 
in the text, together with  VLA observations, and the 2 cm Ryle observations
(filled triangles).}
\label{fig12}
\end{figure}

The resulting model light curves are  shown
in figure \ref{fig12}, together with all observations. 
As can be seen, the model  reproduces the light curves 
at the different wavelengths very well. 
The reproduction of the observed slope
at early times is a
result of 
\begin{equation}
F_{\nu}(\tau_{\rm ssa} \gg 1) \propto R^2 S_{\nu}
 e^{-\tau_{\rm ff}} 
\propto R^2 B^{-1/2} e^{-\tau_{\rm ff}} 
\propto t^{2.5} e^{-\tau_{\rm ff}},
\label{eqwsx}
\end{equation} 
(e.g., Chevalier 1982b). This
is  considerably shallower 
than 
 the pure free-free case discussed by FLC96. Because the magnetic
field is mainly determined by the optically thick part of the
spectrum, equation (\ref{eqwsx}) illustrates the sensitivity of the
derived value of $B$ to the free-free optical depth. 

\begin{figure}  
\plotone{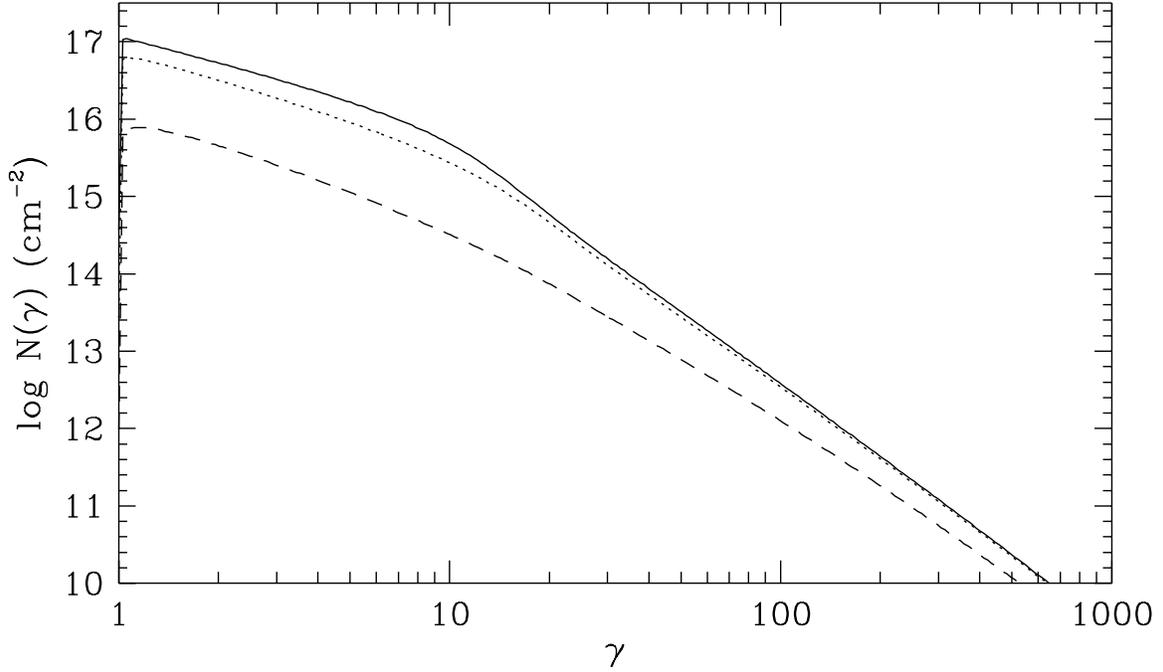}
\caption{
Integrated line-of-sight electron spectra at 10  (solid), 100
(dotted), and 1000 (dashed) days. The flattening of the spectrum caused
by Coulomb losses below $\gamma \sim 10$ is evident at especially 10
and 100 days, while above this energy synchrotron losses prevail. At
10 days synchrotron self-absorption heating gives a weak bump in the
spectrum at $\gamma \sim 10$, corresponding to the energy of the
electrons at which the optical depth is larger than one. 
}
\label{fig13}
\end{figure}
\begin{figure}  
\plotone{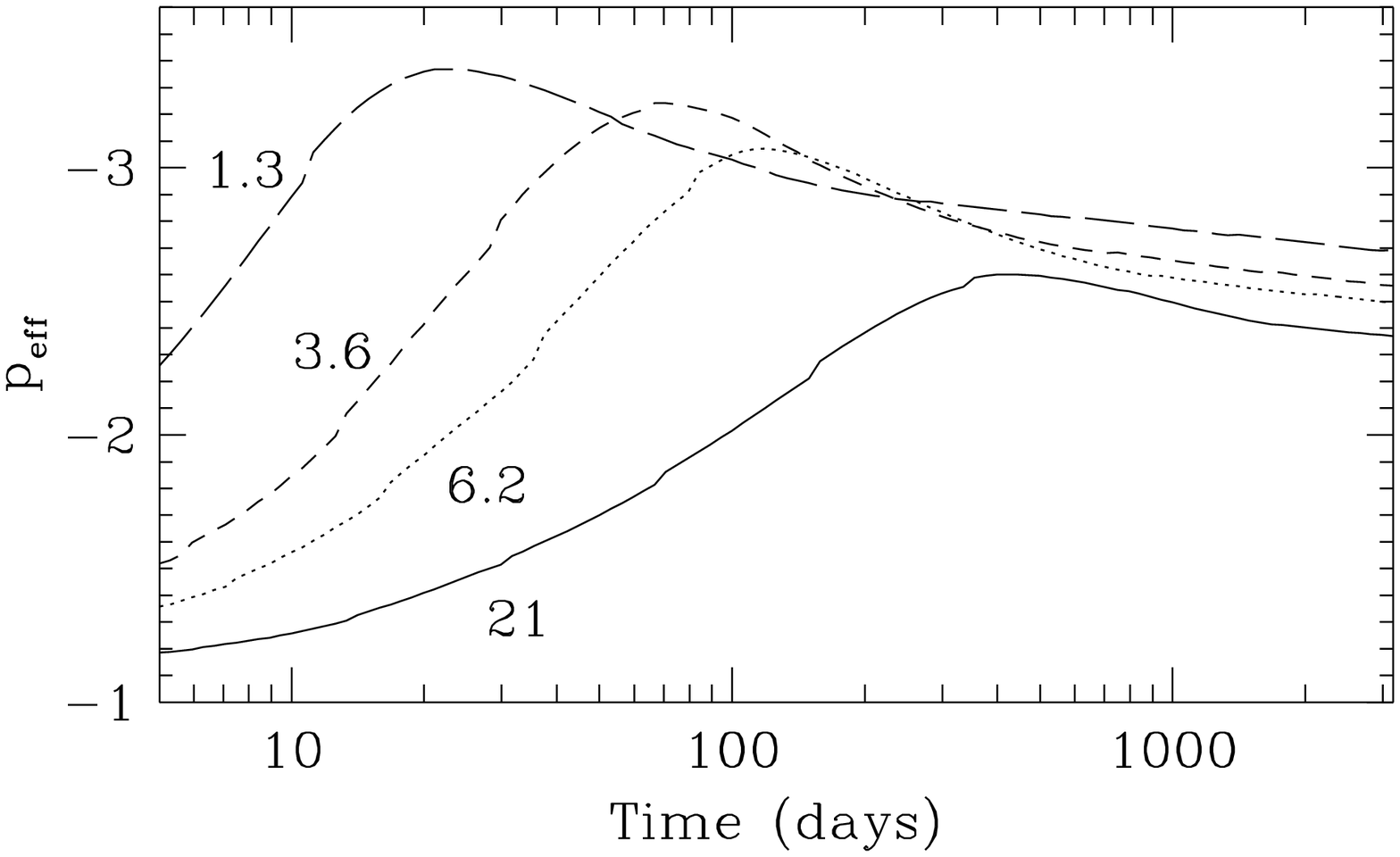}
\caption{The effective spectral index $p_{\rm eff}$ of the electron
distribution at the Lorentz factor ($\gamma = 85 (B \lambda)^{-1/2}$)
 responsible 
for the emission at 1.3 -- 21 cm for the model in figure 12.}
\label{fig14}
\end{figure}

In
figure \ref{fig13} we show the electron distribution at
three different epochs, and in figure \ref{fig14} the
effective spectral index of 
the electron distribution,
$p_{\rm eff} \equiv 2 \alpha + 1$, at the energies responsible for the emission
at the different observed wavelengths. At early times the range is
very large, $p_{\rm eff} = 1.3 -2.5$, while at $t \gsim 1000$ days
the range is smaller at $p_{\rm eff} = 2.4 -
2.8$ (yet still higher than $p_{\rm i}$), showing that cooling is important 
throughout the whole period. The range of $p_{\rm eff}$ even at late times
clearly illustrates the necessity to do a self-consistent calculation of the spectrum,
as has been done here, rather than an analysis based on a constant spectral index as in
section
\ref{sec-ssaff}. The evolution of the electron spectrum
 can be compared directly to figure \ref{fig7} 
where it is seen that for the shortest wavelengths synchrotron cooling
dominates, while at longer wavelengths Coulomb cooling dominates before
$\sim 100$ days, explaining the low value of $p_{\rm eff}$ especially for 21 cm. It is, however,
important to note that the effects of the various cooling processes extend
over a considerable energy range, and the values of $\gamma_{\rm synch}$ and
$\gamma_{\rm Coul}$  are only rough indicators of the cooling effects. 


Synchrotron self-absorption effects on the electron distribution
 are modest and can  be seen only close to
the peak in the spectrum (figure \ref{fig13}). This is the cause of the
crossing of the spectral index curves seen in figure \ref{fig14}.
The reason for the modest self-absorption effects on the electrons is that the
Coulomb timescale  is shorter than the synchrotron timescale for 
$\gamma$ corresponding to $\tau(\nu) \gsim 1$. The electrons are therefore
down-scattered in energy faster than they are redistributed by the 
self-absorption, in agreement with what de Kool \etal (1989) find. 
The optically thick flux is quite 
independent of 
assumptions about spectral index or cooling, which are important when 
 $\tau_{\rm s} \lsim 1$. Only close to the peak in the
spectrum, where $\tau_{\rm s} \sim 1$, can some effects of the self-absorption on the
electron distribution be seen.

Free-free optical depth unity is reached on days 12.6, 26, and 80 for
3.1,  6.2 and 21 cm, respectively. At the shorter wavelengths it never
exceeds  unity because of the high circumstellar temperature. 
The Razin effect is unimportant for the flux at all wavelengths
because of the necessarily high magnetic field (equation \ref {eqhq}).

Both the optically thin luminosity and the optical depth are
determined by the line-of-sight integrated number of relativistic
electrons per energy, $N(\gamma)$, and a qualitative discussion of this
quantity is of interest. For this purpose we  neglect the
self-absorption effects on the electron spectrum, as well as the Razin
suppression. 
If we write equation (\ref{eqnrel}) in terms of the  synchrotron timescale 
$t_{\rm synch} = 6 \pi m_e c /  \sigma_{\rm T} \gamma B^2$, we get 
\begin{eqnarray}
N(\gamma) &=& {dn \over d\gamma} \Delta R_{\rm em} =
{3 \pi  m_e c V n_{\rm  rel} \over 2 \sigma_{\rm T} 
B^2 \gamma_{\rm min}^2} 
\left(
{\gamma \over \gamma_{\rm min}}\right)^{-(p_{\rm i}+1)} \cr
&&\left[1 + 
C(t) + \left({\gamma_{\rm s-C} \over \gamma}\right)^{2}
+ {1.0 \over \gamma} \left(B_{10} \over 30 {\rm ~G}\right)^{-2}
\left({t \over 10~{\rm days}}\right)
\right]^{-1},
\label{eqzq} 
\end{eqnarray}
where 
\begin{equation}
C(t) = 2.49\EE{-2} \left(B_{10} \over 30 {\rm ~G}\right)^{-2} 
\left({L_{\rm Bol}(t) \over 10^{42} \ergs}\right) 
\left({V \over 2\EE4 \kms} \right)^{-2}.
\label{eqat2} 
\end{equation}
The term $C(t)$ gives the ratio of inverse Compton to synchrotron
losses, and can in our case be neglected. The third term in the bracket
gives the Coulomb to synchrotron losses (equation \ref{eqay}), and the 
last, the adiabatic losses. 
When synchrotron losses dominate the factor $\gamma^{-(p_{\rm i}+1)}$ means that
the integrated electron spectrum is  
steepened by one power, while if Coulomb losses dominate it is
flattened by one. 

Using equations (\ref {eqzq}), divided by $4 \pi$, and a delta function approximation for
the total emissivity of a single electron in an isotropic magnetic
field, $j(\gamma, \nu) = c \sigma_T 
B^2 \gamma^2 \delta(\nu - \nu_0 B \gamma^2)/6 \pi$ and 
$j_\nu \Delta R_{\rm em} = \int N(\gamma) / 4 \pi~ j(\gamma,\nu) d\gamma$, 
in equation (\ref{eqa}) with $\bar{\tau}_{\rm ff} \ll 1$, we find  
\begin{eqnarray}
 F_{\nu}(\tau_{\rm ssa} \ll 1) = && {1 \over 4} 
\left({R_{\rm s} \over D}\right)^2 m_e c^2 V
n_{\rm rel} \gamma_{\rm min}^{p_{\rm i}-1}
(\nu_0~B)^{(p_{\rm i}-2)/2} \nu^{-p_{\rm i}/2} \cr
&&\left[1 + C +{\nu_{\rm Coul} \over \nu} 
\left({t \over 10 {~\rm days}}\right)^{-1}  +
\left({\nu_{\rm ad} \over \nu}\right)^{1/2} 
\left({t \over 10 {~\rm days}}\right)^{1/2}
\right]^{-1}
\label{eqalthin} 
\end{eqnarray}
where 
\begin{equation}
\nu_{\rm Coul}=9.32\EE9 \left(B_{10} \over 30 {\rm ~G}\right)^{-1} ~\Mu5~
\left({V \over 2\EE4 \kms}\right)^{-2} ~ {\rm Hz},
\label{eqat3} 
\end{equation}
and
\begin{equation}
\nu_{\rm ad}=1.25\EE8 \left(B_{10} \over 30 {\rm ~G}\right)^{-3} ~ {\rm Hz}.
\label{eqat4} 
\end{equation}
With equations (\ref{eqac}), (\ref{eqebmat}) and (\ref{eqzqq}) we get
\begin{eqnarray}
 F_{\nu}(\tau_{\rm ssa} \ll 1) = && 1.41\EE8 f ~\Mu5 
\left(B_{10} \over 30 {\rm ~G}\right)^{p_{\rm i}/2-1}
\left({V_0 \over 2\EE4 \kms}\right) \cr
&& \left({1.26\EE8 \over \nu}\right)^{p_{\rm i}/2} 
\left({t \over 10 {~\rm days}}\right)^{-(2 \epsilon +1)
(1-m)-(p_{\rm i}-2)/2} \cr
&& \left[1 + C(t) +
{\nu_{\rm Coul} \over \nu}\left({t \over 10 {~\rm days}}\right)^{-1} 
 +
\left({\nu_{\rm ad} \over \nu}\right)^{1/2} 
\left({t \over 10 {~\rm days}}\right)^{1/2}
\right]^{-1} ~ {\rm mJy}.
\label{eqalthin2} 
\end{eqnarray}
 The slow increase in the flux 
of the optically thin 
light curves at 1.3 and 2 cm before 100 days is caused by the decreasing importance of Coulomb losses with time.  
Furthermore, taking $\alpha = d \ln F_\nu / d \ln \nu$ and using $p_{\rm eff} = 
2 \alpha + 1$ gives a good representation of the evolution of the 
electron spectral index at the various wavelengths, as shown in figure 
\ref{fig14}.

After  $\sim 100$ days
the shock velocity decreases according to $m \approx 0.74$. Neglecting the variation of the
bracket in the denominator of equation (\ref{eqalthin2}),
 we get  for $\gsim 100$ days
$ F_{\nu}(\tau_{\rm ssa} \ll 1)
\propto t^{-0.31 -0.52 \epsilon}.
$
The evolution of the optically thin flux is therefore considerably
steeper if $\epsilon=1$, $F_{\nu}(\tau_{\rm ssa} \ll 1) \propto
t^{-0.83}$, compared to $\epsilon=0$. 
Comparing with the observations after $\sim 100$ days there is  much better 
agreement with the $n_{\rm rel} \propto
\rho_{\rm wind}~V^2$ model, compared to the $n_{\rm rel} \propto \rho$
model. The $n_{\rm rel} \propto
\rho_{\rm wind}$ model gives too slow a decline for reasonable values
of
$p_{\rm i}$ and $m$. This is consistent with the conclusions in section 
\ref{sec-nonth}. We also note the insensitivity of the optically thin
flux to the magnetic field when $p_{\rm i}
\approx 2$, $F_{\nu} \propto B^{(p_{\rm i}-2)/2}$, explaining the difficulty 
of separating
the $B \propto R^{-1}$ and  $B \propto t^{-1}$ cases.  
 
 The break 
in the light 
curves at $\sim 100$ days occurs in our models
because of the decrease in velocity, rather than as a result of a 
transition to the adiabatic phase. The rather abrupt change in the light curves, as
well as in  $n_{\rm rel}$ in figure \ref{fig11}, shows that the transition from
constant velocity to a more rapid decrease is fairly sharp, and should
reflect a similarly well-defined flattening in the ejecta structure. 

For our
parameters adiabatic cooling starts to dominate at
$t \sim 1000$ days. However, as can be seen from figure \ref{fig14}, 
the spectral index even at 3000 days
is considerably higher than the input spectral index, $p_{\rm in} =2.1$. 
In the adiabatic phase the optically thin flux  scales as 
\begin{equation}
F_{\nu}(\tau_{\rm ssa} \ll 1) \propto t^{ -(p_{\rm i}-1)/2 - (1+2\epsilon)(1-m)}.
\label{eqwsxy}
\end{equation}

We have also experimented with different slopes of the injection spectrum,  
$p_{\rm i}$. The most sensitive part of the spectrum is not surprisingly 
the epoch later than $\sim 100$ days, when most of the spectrum is 
optically thin. At these epochs the light curves are quite sensitive to 
$p_{\rm i}$, and variations in the range $2.0 \lsim p_{\rm i} \lsim 
2.2$ change the fit to the light curves markedly (as shown by the 
$\chi_{\nu}^2$). 
Our conclusion is therefore that  the spectral index of the
injected electrons should be in the above range, and not be changing in time. 
A value $p_{\rm i} = 2.1$ is remarkably close to what diffusive shock acceleration gives, 
while it is not obvious why other acceleration mechanisms, like turbulent  acceleration
behind the shock, should give the same result.  

Another conclusion is that we can check the validity of 
the approximate equation (\ref{eqnrel}) for the electron spectrum compared to the 
full solution of equation (\ref{eqzqqa}). We find that the difference 
between these approaches is 
small in both the magnetic field and non-thermal electron density. This 
is illustrated by the fact that our approximate analytical solutions 
(\ref{eqwsx}) 
and (\ref{eqalthin2}) accurately describe the numerical light curves.
The results in section \ref{sec-disc} should therefore not be
affected by the additional effects included in this section.

\section{FREE-FREE ABSORPTION, SYNCHROTRON SELF-ABSORPTION, AND THE RAZIN EFFECT FOR RADIO SUPERNOVAE}
\label{sec-freefreecond}

One may  ask under what circumstances free-free absorption,
synchrotron self-absorption, 
or the Razin effect will be important. This is discussed
extensively from the point of view of different supernova categories
by Chevalier (1998). Here we limit ourselves to a few supplemental
remarks.  The radial optical depth to free-free absorption is
\begin{equation}
\tau_{\rm ff}  \approx 4.2~\lambda^2~
\Mu5^{2}~\left({T_{\rm e} \over 10^5 {\rm ~K}}\right)^{-3/2}~
\left({V \over 2\EE4 \kms}\right)^{-3}~
\left({t \over 10 {~\rm days}}\right)^{-3}.
\label{eqtauff3}
\end{equation}
 In the synchrotron cooling domain, and assuming for simplicity that
$p_{\rm i} = 2$, the synchrotron optical depth is
 $\tau_{\rm s}  \approx 1.6\EE4~ f_{\rm rel}~ \Mu5~ (B_{10} / 30 {\rm~ G})^{1/2} (V / 2\EE4 \kms)^{-1}~
 (t / 10 {~\rm days})^{-5/2}
\lambda^{7/2}$.
  From this we note that $\tau_{\rm s}$ in this case is fairly
insensitive to the magnetic field. In the adiabatic regime we 
get $\tau_{\rm s}  
\approx
1.9\EE5~ f_{\rm rel}~ \Mu5~ (B_{10} / 30 {\rm~ G})^{2} (V / 2\EE4 \kms)^{-1} ~ (t / 10 {~\rm
days})^{-3}
\lambda^{3}$. In accordance with our results in section \ref{sec-magn}, we 
further assume that the magnetic field is some fraction
$\phi$ of the equipartition value, $B_{10} \approx 1\EE2 \phi^{1/2} \Mu5$ G.
The ratio of the synchrotron self-absorption and free-free optical depths is then
in the cooling case
\begin{equation}
{\tau_{\rm s} \over  \tau_{\rm ff}}  \approx 6.7\EE3~ f_{\rm rel}~ \phi^{1/4} \Mu5^{-3/4}
\left({T_{\rm e} \over 10^5 {\rm ~K}}\right)^{3/2}~\left({V \over 2\EE4 \kms}
\right)^{2}~\left({t \over 10 {~\rm days}}\right)^{1/2} \lambda^{3/2},
\label{eqtausff1}
\end{equation}
and in the adiabatic case
\begin{equation}
{\tau_{\rm s} \over  \tau_{\rm ff}}  \approx 4.8\EE5~ f_{\rm rel}~ \phi ~
\left({T_{\rm e} \over 10^5 {\rm ~K}}\right)^{3/2}~\left({V \over 2\EE4 \kms}
\right)^{4}~ \lambda.
\label{eqtausff2}
\end{equation} 
With the same
assumptions, the ratio of the synchrotron cooling time to the expansion time is 
\begin{equation}
{t_{\rm synch} \over  t}  \approx 1.1\EE{-2} ~ \phi^{-3/4} ~\Mu5^{-3/2}
 ~\left({t \over 10 {~\rm days}}\right)^{1/2}~ \lambda^{1/2}.
\end{equation}
Unless $\Mu5 \ll 1$ and/or $\phi \ll 1$, synchrotron cooling is likely to be
important.
 
 Equation (\ref{eqtausff1}) shows that in the cooling case 1) 
free-free absorption  decreases in importance with time relative to
synchrotron self-absorption, and 2) that, while insensitive to
the magnetic field, the ratio is highly sensitive to the velocity and
temperature of the circumstellar medium. 

The magnetic field is probably related to the progenitor system, and
may differ 
between normal Type II, Type IIb and Ib supernovae. Because the latter types
are likely to be components in binary systems, this fact may influence the
circumstellar magnetic field. In addition, the turbulent amplification
behind the shock 
may depend on the properties of the shock (e.g.,  whether it is
cooling or not). Because of the lack of knowledge of this quantity it
is fortunate that the optical depth is not very sensitive to the 
magnetic field. 

The circumstellar temperature is  especially sensitive to the expansion
velocity, since this, together with the mass loss rate,
determines the shock temperature, and therefore also the
EUV and soft 
X-ray flux from the supernova (Lundqvist \& Fransson 1988). A higher
expansion velocity leads to a higher shock temperature and a stronger
UV and soft X-ray spectrum from Comptonized radiation, and thereby to
a higher circumstellar temperature. Observationally,
both  SN 1993J and SN 1994I (Filippenko et al. 1995) had higher expansion velocity ($\gsim 2\EE4
\kms$) than SN
1979C and SN 1980K ($\sim 1.2\EE4 \kms$). In addition, the color temperature of the
initial burst of the supernova is higher for a compact progenitor, like for 
a Type Ib/c or
IIb supernova, which increases the heating 
of the circumstellar medium further. The difference in the
supernova properties  is  illustrated in the different circumstellar temperatures
found for SN 1979C and 1980K, compared to SN 1993J. In the former cases Lundqvist \&
Fransson (1988) found a peak temperature of
$\sim 2\EE5$ K, while in the latter case the peak temperature was $\gsim
10^6$ K (FLC96). Even more important, at 100 days the temperature was still
$\sim 2\EE5$ K for SN 1993J, while it was only $(2-3)\EE4$ K for SN
1979C and SN 1980K. 

Therefore, equations (\ref{eqtausff1}) and (\ref{eqtausff2}) show that a higher
expansion velocity increases the importance of synchrotron self-absorption relative to
free-free absorption both directly and through the temperature.

One should note that even if synchrotron self-absorption is larger than
free-free, the rise of the light curve may be dominated by the free-free, since this acts
externally to the synchrotron emitting region. Because $\tau_{\rm ff}$ itself is so
sensitive to the velocity and temperature (equation \ref{eqtauff3}), the sensitivity to the velocity 
and progenitor structure is still present.
The relative importance of synchrotron
self-absorption and free-free absorption may therefore  mainly be a  result of the
higher expansion velocity for the Type IIb and Ib/c supernovae compared to
ordinary Type II supernovae. 

For objects which are dominated by free-free absorption, equation
(\ref{eqtausff1})  gives rise to the question whether 
synchrotron self-absorption
may be important for the underlying spectrum also in these sources. 
Because of the limited number of
observed frequencies this question is difficult to answer, although the above
estimates makes this plausible. 

That a high expansion velocity is indicated for the supernovae showing
synchrotron self-absorption has earlier been argued by Shklovskii
(1985) and more recently by Chevalier (1998). These authors have
discussed especially the radio emission from the Type Ib supernovae, and
conclude that the expansion velocities of these indeed have to be very
high. The sensitivity to the temperature has, however, been discussed little before.

As was illustrated in section \ref{sec-onecomp}, the main problem for the
Razin effect to be important is that it requires a very low magnetic
field. To give a detectable radio emission it then follows that the
column density of relativistic particles has to be unrealistically
high. The high thermal plasma density needed for the Razin effect to
dominate also means that Coulomb losses will become important. We
therefore do not expect the Razin effect to be very important for
supernovae showing an observable radio emission in other than quite
exceptional cases.  

\section{GENERAL CONSEQUENCES FOR VLBI OBSERVATIONS}
\label{sec-vlbi}

Both the dominance of the synchrotron self-absorption and the importance of
electron energy losses have  consequences for the VLBI 
observations. At wavelengths $\lambda \gg \lambda_{\tau_{\rm ssa}=1}$
the intensity will 
be that of a uniform disk, while at  $\lambda \ll \lambda_{\tau_{\rm
ssa}=1}$ limb brightening 
is important (Marscher 1985). For this latter 
situation the full set of equations
(\ref{eqssa})-(\ref{eqssa2}) should be 
used. Free-free absorption with the consequent limb darkening is only
important for the longest wavelengths. 
VLBI observations by Bartel \etal (1994) and Marcaide
\etal (1997)  
have been obtained at 1.35 and 2.01 cm  from day 30 to day 91, 
at 3.56 cm  from day 50 to day 427, and at 6.16 cm from day 91 to day 1304. 
We find that the synchrotron optical depth is larger than unity up to day 
16, 32, 63, and 125 at 1.35, 2.01, 3.56 and 6.16 cm, respectively.
At times earlier than these epochs a uniform disk is therefore to be
preferred as input to the VLBI modeling, while   
a limb brightened shell is more applicable after these epochs.

Synchrotron losses have been shown to be
important throughout the whole period of observations. If 
acceleration  takes
place close to the circumstellar shock in a very thin region, the emitting 
region
will have a wavelength dependent thickness
\begin{equation}
\Delta R_{\em} \approx {V t_{\rm synch} \over 4} \propto V B^{-3/2} 
\lambda^{1/2}
\end{equation}
In this ideal case the thickness would therefore increase with wavelength
 by a factor
of $\sim 40 ~\%$ from 3 cm to 6 cm.
One would also predict that the thickness would increase with time as
the synchrotron 
timescale increases (figure \ref{fig7}).  If the	 acceleration process 
takes place 
in the turbulent
region behind the shock, possibly connected to the contact
discontinuity, a more uniform emissivity would result. As long as synchrotron
cooling dominates, the emissivity is not sensitive to the magnetic
field distribution, because the flux is instead determined by the available energy in
relativistic electrons. The opposite is true if the electrons are adiabatic. 

For the wavelengths where the supernova is optically
thick to synchrotron self-absorption, one expects the polarization to
be low, while in the optically thin phase the polarization can be
high, unless the magnetic field is highly random. Because of the high
degree of spherical symmetry the integrated polarization is
likely to be low. VLBI polarization maps would therefore be
highly interesting, although difficult to obtain in practice. A
few resolution elements over the disk would suffice for this
purpose. 

\section{CONCLUSIONS} 
\label{sec-concl}

Here we summarize our main conclusions:
\begin{enumerate}
\item The light curves are characterized by synchrotron
self-absorption in combination with external free-free absorption. 
While giving an excellent fit to the spectrum, the Razin effect can be 
excluded, based on an unrealistic injection efficiency.
\item The circumstellar density is consistent with a constant mass loss rate and wind velocity,  $\rho_{\rm wind} \propto r^{-2}$.
\item The free-free absorption inferred from the combined synchrotron-self 
absorption plus free-free model  agrees
well with that estimated previously by FLC96.
\item The magnetic field in the synchrotron emitting region can be
reliably determined, and decreases like $B \propto R_{\rm s}^{-1}$, or
alternatively as $B \propto t^{-1}$. The strength argues strongly for turbulent
amplification behind the shock. The energy density is somewhat 
weaker than equipartition, $u_{\rm B}/ u_{\rm therm} \approx 0.14$.
\item Both synchrotron and Coulomb losses are important for the
integrated electron spectrum, which cannot be characterized by a simple power
law.  
\item The injected electron spectrum has a power law slope $p_{\rm i} = 
2.1$, indicating 
that diffusive shock acceleration is operating.
\item The number density of relativistic electrons scales as a fixed
fraction of the 
thermal energy density behind the shock. This may argue for an
injection mechanism which is controlled by the available thermal energy behind
the shock. While the energy density in the relativistic electrons is
 only $\sim 5\EE{-4}$ of the thermal energy density, the ions may have a 
total energy density comparable to equipartition.
\item The light curves, as well as the evolution of the injection rate of
relativistic electrons, reflect the dynamics of the shock wave and therefore also the
ejecta structure. 
\end{enumerate}
\acknowledgments We are grateful to  Lewis Ball,  
Dick McCray, Kurt Weiler and especially Roger Chevalier 
and Peter Lundqvist for many informative and
interesting discussions, and to Kurt Weiler for making the VLA data
publicly available. We are also grateful for many useful 
comments by the referee.
This research was supported by the Swedish Natural Sciences Research Council
and the G\"oran Gustafsson Foundation  for Research in Natural Sciences and
Medicine. Part of this work was done while CF was a visitor at the
Institute
for Theoretical Physics at the University of California at Santa Barbara 
Institute, supported by NSF under Grant No. PHY94-071194.

\vfill

\eject

\newpage

\begin{references}
\reference {} Ball, L., \& Kirk, J.~G. 1992, \apj, 396, L 39
\reference {} Bartel, N., \etal 1994, \nat, 368, 610 
\reference {} Chang, J.~S., \& Cooper, G. 1970, J. Comp. Phys. 6, 1
\reference {} Chevalier, R.~A. 1982a, \apj, 258, 790
\reference {} Chevalier, R.~A. 1982b, \apj, 259, 302
\reference {} Chevalier, R.~A. 1984, Ann. N. Y. Acad. Sci., 422, 215
\reference {} Chevalier, R.~A. 1990, in Supernovae, Ed. A.~G.
Petschek, Berlin, Springer, 91
\reference {} Chevalier, R.~A. 1996,  in 
       Radio Emission from the Stars and the Sun, ed. A. R. Taylor and
J. M. Parades  (Provo: ASP Conf. Series), 125-133.
\reference {} Chevalier, R.~A. 1998, \apj, 499, 810
\reference {} Chevalier, R.~A, \& Blondin, J.~M. 1995,  \apj, 444, 312
\reference {} Chevalier, R.~A., Blondin, J.~M., \& Emmering,
 R.~T. 1992,  \apj,  392, 118
\reference {} Chevalier, R.~A, \& Dwarkadas, V.~V. 1995,  \apj, 452, L45
\reference {} Cohen, R.~J. \etal 1987, \mnras, 225, 491
\reference {} de Kool, M., Begelman, M.~C., \& Sikora, M. 1989, \apj,
337, 66
\reference {} Duffy, P., Ball, L. \& Kirk, J.~G. 1995, \apj, 447, 364
\reference {} Filippenko, A. V., et 
al. 1995,  \apjl, 450, L11
\reference {} Fransson, C. 1994, in Circumstellar Media in the Late Stages of
Stellar Evolution, Eds. R.E.S. Clegg, I.R. Stevens \& W.P.S. Meikle, Cambridge 
University Press, Cambridge, 120
\reference {} Fransson, C. 1998, in preparation
\reference {} Fransson, C., Lundqvist, P. \& Chevalier, R.A. 1996,
\apj, 461, 993
(FLC96)
\reference {}  Freedman, W.~ L. \etal 1994, \apj, 427, 628
\reference {} Jun, B., \& Norman, M.~L. 1996, \apj, 472, 245
\reference {} Lewis, J.~R. \etal 1994, \mnras, 266, L27
\reference {} Lundqvist, P.1994,
Circumstellar Media in the Late Stages of Stellar Evolution
R.~E.~S. Clegg, W.~P.~S. Meikle, \& I.~R. Stevens, CUP, Cambridge, 213
\reference {} Lundqvist, P., \& Fransson, C. 1988 \aap, 192, 221
\reference {} Marscher, A.~P. 1985, in Supernovae as Distance
 Indicators, Ed. N. Bartel, Berlin, Springer, 130
\reference {} Marcaide, J.~M., Alberdi, A., Ros, E., Diamond, P., Schmidt, B., 
Shapiro,
I.~I., Baath, L., Davis, R.~J., de Bruyn, A.~G., El\'osegui, P., Guirado, J.~C.,
Jones, D.~L., Krichbaum, T.~P., Mantovani, F., Preston, R.A., Ratner, M.~I., 
Rius, 
A.,
Rogers, A.~E.~E., Schilizzi, R.T., Trigilio, C.,  
Whitney, A.~R., Witzel, A., \& Zensus, A. 1995,  \nat,  373, 44
\reference {} Marcaide, J.~M., Alberdi, A., Ros, E., Diamond, P.,  
Shapiro, I.~I., Guirado, J.~C.,
Jones, D.~L., Krichbaum, T.~P., Mantovani, F., Preston, R.A., 
Rius, A., Schilizzi, R.~T., Trigilio, C.,  Whitney, A.~R., \& Witzel, A. 1995,
Science, 270, 1475  
\reference {} Marcaide, J.~M., Alberdi, A., Ros, E., Diamond, P., 
Shapiro,
I.~I., Guirado, J.~C., Jones, D.~L., Mantovani, F., Peres-Torres, M.~A., 
Preston, R.A.,  Schilizzi, R.T.,  Sramek, R.~A., Trigilio, G., Van Dyk, S.~D.,
Weiler, K.~W., \&  Whitney, A.~R.
 1997,  \apj, 486, L31
\reference {} McCray, R. 1969, \apj, 156, 329
\reference {} Montes, M. J., Kassim, N. E., Weiler, K. W., Sramek, R. A.
        \& Van Dyk, S. D. 1995,  \iaucirc, 6273
\reference {} Nedoluha, G.~E., \& Bowers, P.~F. 1992, \apj, 392, 249
\reference {} Parker, E.~N. 1958, \apj, 128, 664
\reference {} Pacholczyk, A.~ G. 1970, Radio Astrophysics 
	(San Francisco: W. H. Freeman and Company)
\reference {} Phillips, J. A., \&  Kulkarni, S. R. 1993a,  \iaucirc,    No  .
5775
\reference {} Phillips, J. A., \&  Kulkarni, S. R. 1993b,  \iaucirc,    No  .
5884
\reference {} Pooley, G.~G., \& Green, D.~A. 1993, \mnras, 264, L17
\reference {}  Radford, S., Neri, R., Guilloteau, S., \& Downes, D.,
1993, \iaucirc,    No  . 5768
\reference {}Ramaty, R. \& Lingenfelter, R.~E. 1967,
J. Geophys. Res., 72, 879 
\reference {} Razin, V.~A. 1960, Radiofiz., 3, 584
\reference {} Rees, M.~J. 1967, \mnras, 137, 429
\reference {} Richmond, M.~W., Treffers, R.~R., Filippenko, A.~W., Paik, A.~Y.,
Leibundgut, B., Schulman, E., \& Cox, C.~V. 1994,  \aj, 107, 1022
\reference {} Shigeyama, T., Suzuki, T., Kumagai, S., Nomoto K., Saio, H.,
 \& Yamaoka, H. 1994, \apj, 420, 341
\reference {} Shklovskii, I. S. 1985, SvAL, 11, 105
\reference {} Simon, M. 1969, \apj, 156, 341
\reference {} Slysh, V. I. 1990, SvAL, 16, 339
\reference {} Tsytovich, V.~N. 1951, Vestn. Mosk. Univ. Phys. No. 11, 27
\reference {} Van Dyk, S.~D., Weiler, K.~W., Sramek, R.~A., Rupen, M.~P.,  \&
Panagia, N. 1994,  \apj, 432, L115
\reference {} Weber, E.~J., \& Davis, L.,~Jr. 1967, \apj, 148, 217
\reference {} Weiler, K.~W., Van Dyk, S.~D., Sramek, R.~A., Panagia, N., \& 
 1996, in Supernovae and Supernovae Remnants, eds. R. McCray \&
 Z. Wang, CUP, Cambridge,  283
\reference {} Weiler, K.~W., Montes, M.~J., Van Dyk, S.~D., Sramek, R.~A., Panagia, N., \& 
 1998, in 
Proc. of the Workshop SN 1987A: Ten years after, 
eds. M. Phillips \& N. Suntzeff, in press
\reference {} Young, T.~R., Baron, E., \& Branch, D. 1995, \apj, 449, L51
\end{references}
\end{document}